\documentclass{aa}
\usepackage{epsfig}
\usepackage{txfonts}

\usepackage{natbib}
\usepackage[german,english]{babel}
\begin{document}
   \title{Spectroscopy of clusters in the ESO Distant Cluster Survey
   (EDisCS)\thanks{%
   Based on observations obtained at the ESO Very Large Telescope
   (VLT) as part of the Large Programme 166.A--0162 (the ESO Distant
   Cluster Survey).}
   \thanks{%
   Table \ref{spectrocat} is only available in electronic form at the
   CDS via anoymous ftp to cdsarc.u-strasbg.fr (130.79.125.5) or via
   http://cdsweb.u-strasbg.fr/Abstract.html} }

   \subtitle{Redshifts, velocity dispersions \&
   substructure for 5 clusters}

   \author{C.~Halliday \inst{1}, B.~Milvang-Jensen \inst{2},
   S.~Poirier \inst{3}, B.~M.~Poggianti \inst{1}, P.~Jablonka
   \inst{3}, A.~Arag{\' o}n-Salamanca \inst{4}, R.~P.~Saglia \inst{2},
   G.~De~Lucia \inst{5}, R.~Pell{\' o} \inst{6}, L.~Simard \inst{7},
   D.~I.~Clowe \inst{8,}\inst{9}, G.~Rudnick \inst{5}, J.~J. Dalcanton
   \inst{10}, S.~D.~M.~White \inst{5} \and D.~Zaritsky \inst{9}}


   \offprints{C. Halliday et al.}

   \institute{Osservatorio Astronomico, vicolo dell'Osservatorio 5,
              35122 Padova  \email{halliday/poggianti@pd.astro.it}
\selectlanguage{german}
         \and Max-Planck Institut f{\"u}r extraterrestrische Physik, Giessenbachstrasse, D-85748 Garching, Germany  \email{milvang/saglia@mpe.mpg.de}
\selectlanguage{english}
         \and GEPI, CNRS-UMR8111, Observatoire de Paris, section de Meudon, 5 Place Jules Janssen, F-92195 Meudon Cedex, France \email{Pascale.Jablonka@obspm.fr}
         \and School of Physics and Astronomy, University of Nottingham, University Park, NG7 2RD, United Kingdom  \email{Alfonso.Aragon@nottingham.ac.uk}
\selectlanguage{german}
         \and Max-Planck-Institut f{\"u}r Astrophysik, Karl-Schwarschild-Str. 1, Postfach 1317, D-85741 Garching, Germany  \email{delucia/grudnick/swhite@mpa-garching.mpg.de}
\selectlanguage{english}
         \and Laboratoire d'Astrophysique, UMR 5572, Observatoire Midi-Pyrenees, 14 Avenue E. Belin, 31400 Toulouse, France  \email{Roser.Pello@ast.obs-mip.fr}
         \and Herzberg Institute of Astrophysics, National Research Council of Canada, Victoria, BC V9E 2E7, Canada  \email{Luc.Simard@nrc.ca}
\selectlanguage{german}
         \and Institut f{\"u}r Astrophysik und Extraterrestrische Forschung, Universit{\"a}t Bonn, Auf dem Hugel 71, 53121 Bonn, Germany
\selectlanguage{english}
         \and Steward Observatory, University of Arizona, 933 North Cherry Avenue, Tucson, AZ 85721  \email{clowe/dzaritsky@as.arizona.edu}
         \and Astronomy Department, University of Washington, Box 351580, Seattle, WA 98195  \email{jd@astro.washington.edu}
}
   \authorrunning{C. Halliday et al.}

   \date{Received 17 May 2004; accepted 28 July 2004}

   \abstract{We present spectroscopic observations of galaxies in 4
   clusters at $z = 0.7-0.8$ and in one cluster at $z \sim 0.5$
   obtained with the FORS2 spectrograph on the VLT as part of the ESO
   Distant Cluster Survey (EDisCS), a photometric and spectroscopic
   survey of 20 intermediate to high redshift clusters. We describe
   our target selection, mask design, observation and data reduction
   procedures, using these first 5 clusters to demonstrate how our
   strategies maximise the number of cluster members for which we
   obtain spectroscopy. We present catalogues containing positions,
   I-band magnitudes and spectroscopic redshifts for galaxies in the
   fields of our 5 clusters. These contain 236 cluster members, with
   the number of members per cluster ranging from 30 to 67. Our
   spectroscopic success rate, i.e. the fraction of spectroscopic
   targets which are cluster members, averages 50\% and ranges from
   30\% to 75\%. We use a robust biweight estimator to measure cluster
   velocity dispersions from our spectroscopic redshift samples. We
   also make a first assessment of substructure within our clusters.
   The velocity dispersions range from 400 to 1100 $\,{\rm km}\,{\rm
   s}^{-1}$. Some of the redshift distributions are significantly
   non-Gaussian and we find evidence for significant substructure in
   two clusters, one at $z \sim 0.79$ and the other at $z \sim 0.54$.
   Both have velocity dispersions exceeding 1000 $\,{\rm km}\,{\rm
   s}^{-1}$ but are clearly not fully virialised; their velocity
   dispersions may thus be a poor indicator of their masses. The
   properties of these first 5 EDisCS clusters span a wide range in
   redshift, velocity dispersion, richness and substructure, but are
   representative of the sample as a whole. Spectroscopy for the full
   dataset will allow a comprehensive study of galaxy evolution as a
   function of cluster environment and redshift.

   \keywords{galaxies: clusters: general -- galaxies: distances and
   redshifts -- galaxies: evolution}}

   \maketitle
%

\section{Introduction}

Galaxy clusters are convenient and popular sites for studying
environmental influences on galaxy evolution and the systematic
analysis of large cluster samples is an important step towards a
quantitative understanding of the processes involved. Large area
cluster searches began with the pioneering photographic surveys of
\citet{Abell:1958}, \citet{Zwicky:1968} and \citet{Shectman:1985}.
More recently, surveys for high-redshift clusters
(e.g. \citealt{Gunn_etal:1986}; \citealt{Gioia_etal:1990};
\citealt{Couch_etal:1991}; \citealt{Postman_etal:1996};
\citealt{Gladders_etal:2000}; \citealt{Gonzalez_etal:2001}) have begun
to provide large enough samples to allow an exploration of the
evolution of clusters and the galaxies within them out to significant
lookback times.

Spectroscopic surveys of the full galaxy population of distant galaxy
clusters are few in number. At intermediate redshift they include the
studies of \citet{Couch_etal:1987}, and of the CNOC and MORPHs
collaborations (\citealt{Yee_etal:1996}; \citealt{Balogh_etal:1997};
\citealt{Dressler_etal:1999}; \citealt{Poggianti_etal:1999}), as well
as the studies of individual clusters by van Dokkum, Kelson, Tran and
collaborators (e.g. \citealt{vanDokkum_etal:2000};
\citealt{Tran_etal:2003}). At higher redshift there is the survey by
Postman and collaborators (\citealt{Postman_etal:1998};
\citealt{Postman_etal:2001}) and the study of the single very rich
cluster MS1054-03 by
\citealt{vanDokkum_etal:1999}. Such surveys have contributed
substantially to our understanding of the star formation histories of
cluster galaxies, revealing strong evolution in their average star
formation activity since $z \sim 0.5$, thus over the last 5 Gyr. At
redshifts beyond 0.5, too few clusters have been studied so far to
draw strong conclusions about galaxy evolution and the dependence of
galaxy properties on environment (e.g. van Dokkum et al. 1999, 2000;
Postman et al. 1998, 2001).

Measurements of cluster velocity dispersions are often used to
estimate cluster masses (e.g. \citealt{Fisher_etal:1998};
\citealt{Tran_etal:1999}; \citealt{Borgani_etal:1999};
\citealt{Lubin_etal:2002}). Early measurements of velocity dispersions
(e.g. \citealt{Danese_etal:1980}) usually assumed a Gaussian velocity
distribution. As datasets have increased in quality and size,
significant deviations from Gaussian behaviour have often been found
(e.g. \citealt{Zabludoff_etal:1993}; \citealt{Fisher_etal:1998}).
These are likely associated with the substantial structural
irregularities (usually termed substructure) which are observed in
many clusters (Geller \& Beers 1982; \citealt{Dress_Shect:1988}). Any
statistic used to estimate cluster velocity dispersion should
therefore be robust against outliers and against variations in the
shape of the underlying velocity distribution
(\citealt{Beers_etal:1990}). Cluster velocity dispersion measurements
have been published for only a small number of high redshift clusters
(e.g. \citealt{Tran_etal:1999};
\citealt{Lubin_etal:2002}) and analyses of cluster substructure at
such high redshift are scarce. In this paper we present velocity
dispersions and substructure analyses for 5 of the high redshift
clusters in the ESO Distant Cluster Survey (EDisCS).

The ESO Distant Cluster Survey is a photometric and spectroscopic
survey of 20 clusters selected from the Las Campanas Distant Cluster
Survey (LCDCS) (\citealt{Gonzalez_etal:2001}). Clusters in the LCDCS
were selected to be amongst the most significant surface brightness
peaks in a smoothed drift scan map of 110 square degrees of the
southern sky. One half of our sample was chosen to have high estimated
redshift in the LCDCS ($z_{est} \sim 0.8$) while the other half was
chosen to have intermediate estimated redshift ($z_{est} \sim
0.5$). The apparent reality of these candidates was established before
inclusion in the final EDisCS sample by moderate depth VLT imaging in
two passbands. This allowed verification of the expected
overabundance, colour and magnitude of red sequence galaxies
(\citealt{Gonzalez_etal:2002}).

Our EDisCS observing programme is now complete. It involves deep
optical photometry using FORS2 at the VLT (14 nights), near-IR
photometry with the SOFI instrument at the NTT (20 nights), and
multi-slit spectroscopy using the FORS2 spectrograph at the VLT (22
nights). Our optical photometry comprises V, R and I imaging for the
10 high-redshift clusters and B, V and I imaging for the 10
intermediate-redshift clusters (White et al. 2004, in preparation).
The field of view of the FORS2 instrument is $6.8^{\prime} \times
6.8^{\prime}$. After dithering, the field of view with the maximum
depth of exposure in our imaging data was approximately $6.5^{\prime}
\times 6.5^{\prime}$. Near-IR data in J and K will be presented in
Arag{\' o}n-Salamanca et al. 2004 (in preparation). The field of view
of our SOFI images, taking account of the dithering and overlapping of
exposures, was $6.0^{\prime} \times 4.2^{\prime}$ for our clusters at
z$\sim$0.5 and $5.4^{\prime} \times 4.2^{\prime}$ for our clusters at
z$\sim$0.8: see Arag\'on-Salamanca et al. 2004, in preparation, for
details. This ground-based dataset has been complemented in
February-June 2003 by 80 orbits of HST ACS imaging for our 10
high-redshift clusters (Desai et al. 2004, in preparation).

The EDisCS project aims to investigate galaxy morphology, structure,
masses, stellar populations and star formation histories as a function
of both redshift and cluster environment, and to study cluster mass
and structure. Our spectroscopy allows cluster membership to be
defined using precise redshifts and it constrains the evolutionary
histories of galaxies in dense environments by providing estimates of
(or at least constraints on) star formation rates and histories,
chemical abundances, internal kinematics, and stellar and dynamical
masses. We are studying both early- and late-type galaxy evolution by
measurement of the Fundamental Plane and Tully-Fisher scaling
relations as a function of redshift and environment. The ages and
metallicities of cluster absorption-line galaxies are constrained
using absorption line-strength index measurements and evolutionary
population synthesis models. The star formation histories of cluster
galaxies are being studied using absorption line and emission line
information, as well as broad-band spectral energy distributions and
HST/ACS morphologies.

In this paper we present the procedures we have adopted to carry out
the spectroscopic part of the EDisCS programme, and we give first
results for a set of five clusters. The main focus of this paper is a
presentation of the spectroscopic data for these clusters, one of
which is at $z \sim 0.5$ (Cl~1232.5$-$1250) while the other four are
at $z = 0.7 - 0.8$ (Cl~1040.7$-$1155, Cl~1054.4$-$1146,
Cl~1054.7$-$1245 and Cl~1216.8$-$1201). We will abbreviate these names
as Cl~1232, Cl~1040, Cl~1054$-$11, Cl~1054$-$12 and Cl~1216,
respectively, in the following.
In Section \ref{observe} we describe our target selection strategy,
our mask design and our VLT FORS2 observations. In Section
\ref{reduction} we describe the reduction of these VLT FORS2
observations. In Section \ref{redshifts} we present galaxy redshift
measurements and spectroscopic catalogues, and we assess our
observational completeness and the success of our observing
strategy. In Section \ref{cvddesc} we present velocity dispersion
estimates for our clusters. In Section \ref{substructure} we study
cluster substructure using maps of galaxy positions and also the
Dressler-Shectman (1988) method. A summary is provided in Section
\ref{summary}.


\section{Observations}\label{observe}

\begin{table}
\caption{The EDisCS spectroscopic observing runs}
\label{runinfo}
\begin{tabular}{ccrr}
\hline
Run & Dates & \multicolumn{2}{c}{Number of nights} \\ & (noon-to-noon)
& Total & Usable \\
\hline
Run 1 & 06 Feb 2002 -- 09 Feb 2002 &  3 &  1.6 \\
Run 2 & 11 Mar 2002 -- 19 Mar 2002 &  8 &  6.9 \\
Run 3 & 26 Mar 2003 -- 06 Apr 2003 & 11 & 11.0 \\
Run 4 & 23 Mar 2004 -- 26 Mar 2004 &  3 &  3.0 \\
\hline
\end{tabular}

\end{table}

Spectroscopic observations were completed during the four observing
runs listed in Table~\ref{runinfo}. During our first run we took
relatively short exposures to establish the reality of our clusters,
to measure their redshifts and to tune our photometric redshifts to
allow stricter selection criteria for subsequent runs. During our
second run deep spectroscopic exposures were completed for the 5
EDisCS clusters discussed in this paper, 4 high redshift clusters and
1 intermediate redshift cluster. Short exposures were acquired in
addition in this second run for EDisCS clusters for which data were
not obtained in the first run due to bad weather. During the third
run, one or two additional masks were observed for the 5 clusters
presented in this paper, and long-exposure data were obtained for most
of the other EDisCS clusters. Finally, in the fourth run we completed
our spectroscopic programme for the full sample, obtaining data for
clusters not discussed in this paper.

In section \ref{target} we describe the target selection strategy. In
section \ref{mask} we describe the VLT FORS2 MXU mask preparation. In
section \ref{logs} we describe the observations themselves.

\subsection{Target Selection Strategy}\label{target}

When selecting the spectroscopic targets, the main aim was to produce
a well-defined sample of cluster galaxies that was as statistically
complete as possible. Given the available imaging data, selection in
the I-band was the obvious choice: such red-band selection reduces
biases towards star-forming or unextincted galaxies. Our parent sample
consisted of all objects with $18.6\le I \le 22$ for the $z \sim 0.5$
cluster candidates and $19.5\le I \le 23$ for the $z \sim 0.8$ ones.
Since we are completing spectroscopy using $1^{\prime\prime}$ slits,
for our target selection we used $1^{\prime\prime}$ radius aperture
magnitudes. The bright magnitude cut was imposed to reject galaxies
that are too bright to be cluster members at the expected redshifts of
our clusters. These bright magnitude limits are $\sim 1\,$mag brighter
than brightest cluster galaxies at $z \sim 0.5$ and $z \sim 0.8$
respectively (\citealt{Aragon-Salamanca_etal:1998}), and are thus very
conservative.

Such straightforward I-band selection produces a sample where the
selection criteria are easy to model and to reproduce, but where the
contamination by non-cluster galaxies and by stars is quite high. For
example, for the extremely rich $z = 0.83$ cluster MS~1054-03, Keck
spectroscopy for an I-band selected sample found that only $\simeq
40$\% of the galaxies with measured redshifts were actually cluster
members \citep{vanDokkum_etal:2000}. Our optically-selected clusters
have a range of richnesses, so we would expect our success rate for a
purely I-band selected sample to be substantially worse than this in
most cases. To avoid wasting telescope time, we decided to use the
available imaging information to impose additional selection
criteria. The aim is to reject a significant fraction of non-members
while keeping the sample of cluster galaxies as close as possible to
an I-band selected one. Since completeness of our cluster samples is
our main concern, our rejection criteria were chosen to be
conservative.

Our target rejection criteria were based primarily on photometric
redshifts obtained with a modified version of the code {\tt
hyperz}\footnote{http://webast.obs-mip.fr/hyperz/} (Bolzonella et
al. 2000). Photometric redshifts have proven to be quite reliable when
a reasonably-complete set of photometry, including near-infrared
bands, is available. When tested against spectroscopic redshifts, the
typical errors in the photometric redshifts are $\simeq0.1$
(\citealt{Bolzonella_etal:2000}; Pell\'o et al. 2004, in
preparation). The {\tt hyperz} code is based on a standard SED fitting
minimization procedure, broad-band filters allowing the detection of
strong spectral features, such as the $4000$\,\AA\ break, the Lyman
break or strong emission lines. The template library used includes 6
evolutionary synthetic SEDs computed using the Bruzual \& Charlot code
(Bruzual \& Charlot 1993), matching the observed colours of local
field galaxies at z$\sim$ 0, from E to Im type, in addition to a set
of empirical SEDs compiled by Coleman, Wu and Weedman (1980), and 2
starburst galaxies from the Kinney et al. (1996) template library (SB1
and SB2). The internal reddening law is taken from Calzetti (2000),
and considered as a free parameter with $A_V$ ranging between 0 and
1.5 magnitudes (E(B-V) between 0 and $\sim$ 0.37 mags). For each
cluster field a correction for galactic extinction was included using
the value of E(B-V) derived from Schlegel et al. (1998) for the
cluster centre. Galactic E(B-V) corrections in the EDisCS fields
typically range between 0.03 and 0.08 magnitudes. The reader is
referred to the reference paper (Pell\'o et al. 2004, in preparation)
for a more detailed description.

Photometric redshift accuracy depends on the wavelength domain covered
by the photometric SEDs, i.e. the filter set. The set of filters used
by EDisCS was designed to cover the relevant wavelength domain in the
restframe of mid z$\sim$0.5 and high z$\sim$0.8 clusters, in
particular to bracket the 4000\,\AA\ break within the relevant
redshift intervals. As explained in White et al. (2004), deep optical
photometry was obtained using FORS2 at the VLT, in BVI and VRI bands
respectively for our mid-$z$ and the high-$z$ clusters. In addition,
deep near-IR images were obtained with SOFI at NTT, in K' and JK'
bands respectively for our mid-$z$ and the high-$z$ clusters (see
Arag\'on-Salamanca et al. 2004, in preparation for details). These
optical and near-IR band data were used to derive our photometric
redshifts.

After imposing the magnitude cuts described above, targets were
selected if their photometric redshift obeyed $z_{\rm clus}-0.2\le
z_{\rm phot}\le z_{\rm clus}+0.2$, where $z_{\rm clus}$ was the
estimated cluster redshift. The cluster redshift was estimated,
initially, from the peak in the photometric redshift histograms (see
Pell\'o et al. 2004, in preparation). After our first short-exposure
masks were observed and analysed, cluster redshifts based on our
spectroscopy were used instead.

As an additional measure to ensure that no likely cluster members were
dropped from the sample, we took into account that the photometric
redshift code provides not only a ``best guess'' photometric redshift,
but also a redshift probability distribution for each object. Objects
with $z_{\rm phot}$ outside the $z_{\rm clus}\pm0.2$ interval were
selected for spectroscopy if the $\chi^2$-based probability of the
best fit provided by the photometric redshift code at the cluster
redshift was larger than 50\%. Because the redshift probability
distributions were reasonably peaked for most objects, this only
increased the number of selected targets by 2--3\%. For this reason,
the exact value of the probability threshold for selection had very
little effect on our target list (see Pell\'o et al. 2004, in
preparation for details).

Finally, the photometric redshift code also provides some information
to help with star-galaxy separation. For each object, the code
computes two independent values, $N_{\rm G}$ and $N_{\rm *}$, based on
how well the photometry corresponds to a galactic or a stellar SED
(Pell\'o et al. 2004). $N_{\rm G}=0$ means ``the object is almost
certainly not a galaxy''; $N_{\rm G}=1$ ``the object could be a
galaxy''; and $N_{\rm G}=2$ ``the object is almost certainly a
galaxy''. Similarly, $N_{\rm *}$ can have values 0, 1 and 2 with
similar meanings but for stars.  Objects with $N_{\rm G}=0$ and
$N_{\rm *}\ge 1$ were excluded from the spectroscopic sample as very
probable stars. As a safety precaution, we have checked that the vast
majority (at least $\simeq95$\%) of the objects rejected as probable
stars based on their photometry have stellar PSFs, while virtually all
the rest are blended objects. Moreover, each spectroscopic mask had
slits placed on several stars (typically 3--5) for acquisition
purposes. These stars were selected from the objects photometrically
classified as very probable stars. There were 52 such objects in the
masks of the five clusters discussed in this paper, and the
spectroscopy confirmed all of them as stars.

Indeed, the photometric criteria used to remove stars from the target
list is more conservative than using the PSF alone: a significant
number of objects with stellar PSFs remained in the target list in the
2002 observing runs, and they were confirmed as stars by the
spectroscopy. To further reduce our stellar contamination, in the 2003
and 2004 observing runs we decided to reject in addition objects which
have clear stellar PSFs in the I-band images. This decision was based
on the excellent image quality of our I-band data (median seeing
$0.6^{\prime\prime}$, cf. White et al.\ 2004), and on the experience
gained in the 2002 runs. For each cluster, a set of 20--30 bona-fide
stars was selected, their ellipticities and ${\rm FWHM}$s measured,
and the mean ($\langle {\rm FWHM}\rangle$) and standard deviation of
the ${\rm FWHM}$s ($\sigma_{\rm FWHM}$) determined. Then, objects were
removed from the target list of that cluster if their ${\rm FWHM}$ was
less than $\langle {\rm FWHM}\rangle+2\sigma_{\rm FWHM}$ and their
ellipticities were less than 0.1. Typically, $\langle {\rm
FWHM}\rangle\simeq0.6^{\prime\prime}$ and $\sigma_{\rm FWHM}\simeq
0.05^{\prime\prime}$. This procedure removed only $\sim5$\% of the
candidate targets. Note that the results presented in this paper are
based mostly on spectroscopy obtained in 2002, and thus this
additional selection criterion was not applied there. Had we applied
it, out of the 231 spectroscopically-confirmed cluster members, not a
single one would have been rejected on the basis of its PSF. Indeed,
not a single object with spectroscopically-confirmed redshift larger
than $\simeq0.0$ would have been rejected on the basis of its PSF.

In addition to these photometric criteria, the mask design imposes
geometric constraints in order to avoid overlapping spectra and to
ensure the necessary wavelength coverage (see next section). In the
mask design process, it was sometimes impossible to find targets that
fulfilled all the selection criteria at a given mask location. In such
cases, in order to fill the masks, slits were placed on an ad hoc
basis on objects that had not been selected as cluster galaxy
candidates, and therefore were expected to be field galaxies based on
their photometric redshifts. A test of how well our selection worked
is provided by the number of such photometrically-selected field
galaxies that turned out to be cluster members. For the five clusters
discussed in this paper, redshifts were obtained for 65 such galaxies,
and none of them turned out to be a cluster member. Thus we are
reasonably confident that our selection criteria have excluded, at
most, 1-2\% of real cluster galaxies above our I-band limit. Once the
spectroscopy for our complete EDisCS data set is analysed, we will
have tighter constraints on this number.

A measure of the effectiveness of our selection procedure is that our
success rate, measured as the fraction of spectroscopic targets that
turned out to be cluster members, is, on average, $\simeq 50$\% for
the five clusters described in this paper. If we had used a pure
I-band selection, the success rate would have been under 30\%, and our
use of 8-m telescope time substantially less efficient. In summary, we
are quite confident that our target selection process yielded a
cluster galaxy sample that is very close to a purely I-band selected
one, but with almost twice as many spectroscopically-confirmed cluster
members as would have been the case if we had not made use of
photo-$z$ rejection of non-members.

\subsection{Automatic Creation of the Masks}\label{mask}

We developed a program to automatically place MXU mask slits on
targeted galaxies \citep{Poirier:2004}.

Mask design was performed using our EDisCS I-band images, since these
best correspond to the wavelength domain of the grism 600RI+19 which
we had chosen for our spectroscopy. We fixed the right ascension axis
to be parallel to the X axis of the detectors. The centre of the mask
was chosen in order to maximize the number of galaxies to be
observed. For the 2002 observations we matched the spectroscopic field
centre with the position of the brightest cluster galaxy (BCG) in X
and the center of the photometric image in Y. For 2003, we had to take
into account the change of FORS2 CCDs and in particular a 5\farcs5 gap
between the two new detectors. We allowed a 10 to 20$^{\prime\prime}$
shift along the Y axis.

Reference stars were selected to aid the calculation of shifts in RA,
Dec and rotation required to precisely align each MXU mask on the
sky. Seven reference stars were chosen for each mask to have 17.5 mag
$<$ I $<$ 19.9 mag and to be uniformly distributed across the field of
view. Short ($5^{\prime}$) slits were placed on 2-3 stars to help
optimise the final mask alignment and to allow seeing to be measured
during observations.

Our science aims required the observation of a minimum spectral region
including the [O{\sc II}]$\lambda$3727 and H$\delta\lambda$4101
spectral lines for the cluster galaxy candidates. This restframe
domain was shifted for each cluster, according to its photometric
redshift for the preparation of our first February 2002 spectroscopic
run, or using the spectroscopic cluster redshift for the preparation
of our long-exposure masks.

The slit width was 1$^{\prime\prime}$. We selected slit lengths of
1$^{\prime\prime}$ for cluster galaxy candidates, 6$^{\prime\prime}$
for field galaxy candidates, and 5$^{\prime\prime}$ for star reference
slits. For the cluster BCG galaxy a 10$^{\prime\prime}$ slit was not
sufficiently long and we chose a slit length equal to
10$\times$FWHM. We fixed a minimum separation between slits of
1$^{\prime\prime}$ to avoid any superposition of spectra. The galaxies
were centered in the slits when possible, otherwise at least
3$^{\prime\prime}$ away from the slit edges to ensure good sky
substraction.

In 2002, slits on cluster galaxy candidates were aligned with the
galaxy major axis when {\tt hyperz} indicated a spiral type spectral
energy distribution. The position angle provided by SExtractor was
used as long as it did not exceed 45 degrees. A vertical slit was
selected for all remaining galaxies. After reduction of the 2002 runs,
we identified a number of emission line galaxies which had been
classified as ellipticals by {\tt hyperz}. In subsequent runs slits
were aligned with the major axis for all galaxies provided that the
slit inclination angle did not exceed 45 degrees.

Galaxies were selected by order of priority: 1) BCG or the closest
possible galaxy; 2) confirmed cluster galaxy, or target for which the
redshift determination was still ambiguous in the short exposure
spectra; 3) cluster galaxy candidates, based on the criteria presented
in Section \ref{target}; 4) star slits; 5) field galaxy
candidates. For each of the categories 1 to 3, the spectroscopic
target selection was completed so that the brightest cluster member
candidates in the considered region of the mask were selected first.

Once all possible primary targets had been exhausted, we filled masks
with 6$^{\prime\prime}$ slits placed preferentially on cluster galaxy
candidates, with no constraint on the wavelength range. Finally a
visual inspection of all masks was performed.

%
\subsection{Observations}\label{logs}
Spectroscopic data were acquired during four observing runs using the
MXU multi-object mask facility of the FORS2 spectrograph mounted on
the VLT Yepun UT4 telescope, ESO Paranal, as described in Tables~1
and~3. Details of our observational setup are provided in
Table~\ref{chip}.

\begin{table}
\caption{The MXU multi-slit instrumental set-up} 
\label{chip}
\begin{tabular}{lcc}
\hline
 Run                        & run 1 + 2                       & run 3 + 4  \\
\hline                      
 Telescope                  & \multicolumn{2}{c}{VLT--UT4}               \\
 Instrument                 & \multicolumn{2}{c}{FORS2}                  \\
 Mode                       & \multicolumn{2}{c}{MXU (i.e.\ mask multi-slit)} \\
 Grism                      & \multicolumn{2}{c}{GRIS$\_$600RI+19}       \\
 Filter                     & \multicolumn{2}{c}{GG435}                  \\
 Spectral resolution        & \multicolumn{2}{c}{6.6$\,${\AA}  FWHM }    \\
 Detector(s)                & Site                 & 2 $\times$ MIT \\
 Detector size              & $2048 \times 2048$   & $2 \times (2048 \times 1024)^{\rm a}$ \\
 Pixel Size                 & 24$\mu$m x 24$\mu$m  & 30$\mu$m x 30$\mu$m$^{\rm a}$ \\
 Readout Noise              & 5.41~e$^{-}$         & 3.15~e$^{-}$        \\
 Conversion factor          & 1.91~e$^{-}$/ADU     & 0.70~e$^{-}$/ADU    \\
 MXU slit width             & 1\farcs0             & 1\farcs0            \\
 Spatial scale              & 0\farcs20/pixel      & 0\farcs25/pixel$^{\rm a}$     \\
 Dispersion                 & 1.32$\,${\AA}/pixel  & 1.66$\,${\AA}/pixel$^{\rm a}$ \\ 
 Wavelength range$^{\rm b}$ & 5300--8000$\,${\AA}  & 5120--8450$\,${\AA}           \\
\hline
\end{tabular}

\vspace*{1pt}

$^{\rm a}$~After binning the CCDs by $2 \times 2$, which is the default.\\
$^{\rm b}$~For a slit at the centre of the field of view.

\end{table}

\begin{table*}
\caption{Observing log for the four EDisCS spectroscopic observing runs}
\label{obslog:new_v2}

{\tabcolsep=6.00pt
\begin{tabular}{lrrrrrrrrrrrl}
\hline
                                       & \multicolumn{11}{c}{\rule{16.5em}{0.4pt}Exposure time [s]\rule{16.5em}{0.4pt}} & \\
                                       & \multicolumn{1}{c}{\rule{0.0em}{0.4pt}Run 1\rule{0.0em}{0.4pt}}
                                       & \multicolumn{3}{c}{\rule{3.8em}{0.4pt}Run 2\rule{3.8em}{0.4pt}}
                                       & \multicolumn{4}{c}{\rule{5.5em}{0.4pt}Run 3\rule{5.5em}{0.4pt}} 
                                       & \multicolumn{3}{c}{\rule{3.8em}{0.4pt}Run 4\rule{3.8em}{0.4pt}} 
                                       & \\
Cluster                                & \multicolumn{1}{c}{m1}
                                       & \multicolumn{1}{c}{m2}
                                       & \multicolumn{1}{c}{m3}
                                       & \multicolumn{1}{c}{m4}
                                       & \multicolumn{1}{c}{m5}
                                       & \multicolumn{1}{c}{m6}
                                       & \multicolumn{1}{c}{m7}
                                       & \multicolumn{1}{c}{m8} 
                                       & \multicolumn{1}{c}{m9} 
                                       & \multicolumn{1}{c}{m10} 
                                       & \multicolumn{1}{c}{m11} 
                                       & Repeats \\
\hline

\multicolumn{13}{l}{Mid-$z$ clusters:} \\
Cl 1018.8$-$1211                       &  1800 & ~~~~~ &  1500 & ~~~~~ &  5400 &  5400 &  7200 & ~~~~~ & ~~~~~ & ~~~~~ & ~~~~~ & m1--m5; m3--m6 \\
Cl 1059.2$-$1253                       &       &  1800 &       &       &  5400 &  7200 &  7200 &       &       &       &       & m2--m5 \\
Cl 1119.3$-$1129                       &       &  1800 &       &       &       &       &       &       &  3600 &  3600 &  3600 & m2--m9 \\
Cl 1202.7$-$1224                       &       &  1800 &       &       &       &       &       &       &  3600 &  3600 &  3600 & m2--m9 \\
Cl 1232.5$-$1250$^\ast$                &       &  7200 &  6360 &  7200 &  7200 &       &       &       &       &       &       &        \\
Cl 1238.5$-$1144                       &       &  1800 &       &       &       &       &       &       &  1200 &       &       & m2--m9 \\
Cl 1301.7$-$1139                       &       &  1800 &       &       &       &       &       &       &  3600 &  3600 &  3600 & m2--m9 \\
Cl 1353.0$-$1137                       &       &  2400 &       &       &       &       &       &       &  3600 &  3600 &  3600 & m2--m9 \\
Cl 1411.1$-$1148                       &       &  1500 &       &       & 10800 &  7200 &  9000 &       &       &       &       & m2--m5 \\
Cl 1420.3$-$1236                       &       &  1800 &       &       &       &       &       &       &  3600 &  3600 &  3600 & m2--m9 \\[4pt]

\multicolumn{13}{l}{High-$z$ clusters:} \\
Cl 1037.9$-$1243                       &  3600 &       &       &       &  7200 & 10800 & 10800 & 10800 & ~~~~~ & ~~~~~ & ~~~~~ & m1--m5 \\
Cl 1040.7$-$1155$^\ast$                &  3600 & 10800 & 14400 & 14400 & 14400 & 14400 & ~~~~~ & ~~~~~ &       &       &       & m1--m2 \\
Cl 1054.4$-$1146$^\ast$                &  3600 & 10800 & 14400 & 14400 & 14400 &       &       &       &       &       &       & m1--m2 \\
Cl 1054.7$-$1245$^\ast$                &  3600 & 10800 & 14400 & 14400 & 14400 &       &       &       &       &       &       & m1--m2 \\
Cl 1103.7$-$1245                       &  3600 &       &       &       & 10800 & 10800 &       &       & 14400 & 14400 &       & m1--m5 \\
Cl 1122.9$-$1136                       &  3600 &       &       &       &       &       &       &       &       &       &       &        \\
Cl 1138.2$-$1133                       &  3600 &       &       &       &  3600 &  7200 &  7200 &  7200 &       &       &       & m1--m5 \\
Cl 1216.8$-$1201$^\ast$                &  3600 & 10800 & 14400 & 14400 & 14400 &       &       &       &       &       &       & m1--m2 \\
Cl 1227.9$-$1138                       &  3600 &       &       &       & 10800 & 14400 & 14400 & 14400 &       &       &       & m1--m5 \\
Cl 1354.2$-$1230                       &  1800 &       &       &       & 12600 & 14100 & 15000 & 17400 &       &       &       & m1--m5 \\

\hline
\end{tabular}
}

\vspace*{1pt}

Notes --- 
m1, m2, \ldots, m11 denotes mask~1, mask 2, \ldots, mask 11.
The observing run can be deduced from the mask number.
A typical observing strategy was to take a short exposure on a cluster
in one run, followed in a subsequent run by a longer exposure
in which only the spectroscopically confirmed cluster members from
the short mask were repeated.
Such mask pairs are listed in the column `Repeats'.
The five clusters marked `$\ast$' are the main focus of this paper.
These are the clusters for which long exposures were obtained in run~2.

\end{table*}


The observing log is given in Table~\ref{obslog:new_v2}.
For each of the 20 EDisCS clusters the total exposure time for each mask
observed in the four observing runs is listed.
The individual exposures were typically 1800$\,$s.
The masks are labelled in a simple way:
masks used in run~1 are called mask~1,
masks used in run~2 are called mask~2, 3 and 4,
masks used in run~3 are called mask~5, 6, 7 and 8, and
masks used in run~4 are called mask~9, 10 and 11.
Since not all clusters were observed in run~1, not all
clusters have a `mask~1', et cetera.

During run~1 relatively short exposures (typically 1~hour) were
completed for the 10 clusters of our high-$z$ sample and for one
cluster in our intermediate sample. Time lost to bad weather meant
that short exposures could not be obtained for the remaining mid-$z$
clusters in this run. For the data acquired, galaxy redshifts were
measured when possible. Masks for our run 2 were designed to target
established cluster members and additional candidate cluster members
for which short exposures were not obtained or a redshift could not be
measured due to insufficient signal-to-noise.

In run~2 deep exposures were obtained for the 5 clusters that are the
main focus of this paper: Cl~1232 from the mid-$z$ sample and Cl~1040,
Cl~1054$-$11, Cl~1054$-$12 and Cl~1216 from the high-$z$ sample. The
typical exposure time per mask was 2~hours at mid-$z$ and 4~hours at
high-$z$. In addition short exposures were obtained for a further 8
clusters from the mid-$z$ sample. Seeing was in general subarcsecond
and typically varied between 0\farcs7 and 1\farcs7.

In run~3 long exposures were obtained for an additional 1--2 masks for
the 5 clusters that already had long masks from run~2. This completed
the spectroscopic data for these clusters. During run~3 long exposures
were obtained for 8 additional clusters, with 2--4 masks per cluster
totaling 27 masks. In run~4 the planned long exposures on the high-$z$
sample were completed, and semi-long exposures (1~hour per mask) were
carried out for clusters from the mid-$z$ sample. These data will be
presented in detail in forthcoming papers.

The calibration frames obtained in the 4 runs were as follows. During
the bright part of twilight sky flat-field exposures were acquired.
During the darker part of twilight calibration stars were observed.
Hot stars (spectral type late O to early B) for the telluric
absorption correction were observed with a longslit of width
$1\farcs0$. Spectrophotometric standard stars for the flux calibration
were observed using the FORS2 MOS mode. For our purpose this provided
a short `longslit' of width $5\farcs0$ that could be placed anywhere
in the field of view. We completed exposures at the extreme left, the
centre and the extreme right of the field of view, bracketing the
range of slit positions and hence wavelength ranges of our MXU galaxy
slits.

Bias, screen flat-field and He+Ar+Ne+HgCd comparison arc-lamp
exposures were taken by an automatic batch procedure executed after
each night of observation. 5 biases, 6 screen flat-field exposures and
a single arc-lamp exposure were obtained for each mask and set of
stellar observations.


\section{Data reduction}\label{reduction}

Data reduction was completed primarily using IRAF\footnote{%
IRAF is distributed by the National Optical Astronomy Observatories,
which are operated by the Association of Universities for Research in
Astronomy, Inc., under cooperative agreement with the National Science
Foundation.}.
Specialised algorithms developed for the reduction of similar FORS2
MXU data \citep{Milvang-Jensen:2003} were adopted for several of the
reduction steps.
In the reduction of the data from run~3, the images from the two CCDs
(cf.\ Table~\ref{chip}) were reduced separately. We describe the data
reduction of spectra acquired in runs~1 and 2, and for data from run~3
for the five clusters that are the principle focus of this paper.

Bias subtraction was completed by subtracting a combined bias image.
For the run~3 data an overscan level correction was made. Following
this cosmic ray hits (hereafter `cosmics') were identified. Different
approaches were taken depending on the number of consecutive exposures
obtained for a given mask.
For the `long masks' typically 4 or 8 exposures of 30$\,$min were
available. In this case, cosmics were identified using a method
developed by \citet{Milvang-Jensen:2003} described in detail in
Appendix \ref{cosmic}.

Single exposures were treated for cosmics using a median filter of
dimensions 5 pixels along the dispersion axis and 1 pixel along the
spatial axis for data acquired in run~1. Double exposures acquired in
run~1 were treated by comparing each frame and taking the minimum
value of each corresponding pixel value as that of the output
frame. Data obtained in run~2 with 1-3 successive exposures were
processed using the LAcosmic algorithm \citep{vanDokkum:2001}; a very
conservative choice of rejection level was made to prevent removal of
sky-lines; some small cosmics remained and were treated interactively
using the IRAF task {\sc cosmicrays}.

The spectra in the upper part of the field curve like a `U' while the
spectra in the lower part of the field curve like an upside-down `U'.
The maximum peak to peak effect was 1.2$''$ (i.e.\ 6$\,$px for the
2002 data). This geometrical distortion (S--distortion) was mapped
using the edges of the individual `spectra' in the screen flats.
Specifically, the screen flats were convolved with a $[-1,1]$ kernel
after which the absolute value was taken. This produced an image in
which the edges of the spectra show up as positive features (the
convolution is equivalent to moving the image downwards by 1~pixel and
subtracting it from itself.) These features were then used to map the
distortion. All frames (science, flat fields and arcs) were then
corrected for the distortion. This created spectra parallel to the
X axis.

For each mask the combined screen flat exposures were used to
construct two types of flat fields: a pixel-to-pixel flat field and a
flat field in which no normalisation was made in the wavelength
direction. The pixel-to-pixel flat was constructed by normalising each
row using a 20-piece cubic spline, i.e.\ by removing the wavelength
dependence caused by the combined effect of the SED of the screen flat
lamp and the transmission function of the system (mirrors, grism,
CCD). The pixel-to-pixel flat field was applied to the corresponding
science frame. In this way the original counts are preserved, allowing
the photon noise to be readily computed.
The second type of flat field was constructed as follows. The combined
screen flat image for the given mask was cut up and each slit spectrum
wavelength calibrated. The slits in the masks would be at different
$x$-positions, and the slit closest to the centre was located. Only
the overall level in the corresponding screen flat spectrum was
normalised, leaving the wavelength dependence intact. The
normalisation constant was chosen to give a level of one at a
reference wavelength of 6780$\,${\AA}, which is central wavelength of
the grism according to the manual. This normalisation constant was
used to normalise all other screen flat spectra for the mask. These
flat field spectra were used to create a second set of science frames
required for the flux-calibration of science spectra (cf.\ below).

Since the frames had been geometrically rectified, the individual
slitlet spectra within each CCD frame were identified by specifying
the Y coordinate of the lower and upper edges of the spectra
positions. The slitlet edges were automatically defined as the points
where the intensity had dropped to 95\% of the maximum level within
the given slitlet screen flat spectrum. Both science and arc frames
were cut. From this point on, the data reduction was identical to the
reduction of longslit data.

For each slitlet a 2D wavelength calibration was established and
applied using the corresponding cut arc spectrum and standard IRAF
tasks. The rms of the fits was typically 0.05$\,${\AA}. Typically
30--40 arc lines well distributed over the wavelength range were used.

Sky-subtraction was performed by first creating one-dimensional
spatial profiles for each cut wavelength-calibrated science spectrum.
Each profile was examined and suitable intervals along the spatial
axis for which a fit to the sky level were decided. The sky was
modelled as a constant at each pixel in the wavelength direction.

One-dimensional spectra were extracted using the IRAF task {\sc
apall}. The extraction aperture was chosen to be the FWHM of the
spatial profile and within each aperture the flux was summed.

The longslit spectra of the hot stars were reduced following a similar
procedure to that of the MXU data. The issue of the two types of flat
fields is irrelevant here, since the extracted 1D spectra were
normalised to unity in the continuum. In the blue part of the
spectrum, where these stars have intrinsic spectral features, the
level was set to unity. The resulting telluric absorption correction
spectrum was scaled and applied to the galaxy spectra.

As mentioned, slit spectra of the spectrophotometric standard stars
were taken at three positions: right, centre and left, covering three
wavelength ranges: extreme blue, central and extreme red,
respectively. The central spectrum has a substantial overlap with the
two other spectra. However, when only the pixel-to-pixel flat field
was applied, within the overlap regions the spectra had quite
different shapes. We attributed this to the grism having a spectral
response that depends on the position (angle) within the field of
view. After the three spectra were divided by the screen flat field in
which spectral shape was left in (cf.\ above), the spectra {\em did}
match up and could be combined to form a single spectrum covering a
wavelength range bracketing that for all galaxy spectra. From this
spectrum the sensitivity function was derived and applied to the
galaxy spectra that had been reduced using the same type of flat
field. We thus use the fact that the position dependent spectral
response of the grism is recorded both in the individual galaxy and
star spectra and in the corresponding screen flat spectra, and when
the former is divided by the latter, the spectral response of the
grism cancels out, as long as the SED of the lamp stays constant.


   \begin{figure*} 
   \centering
   \includegraphics[width=10.0cm,angle=270]{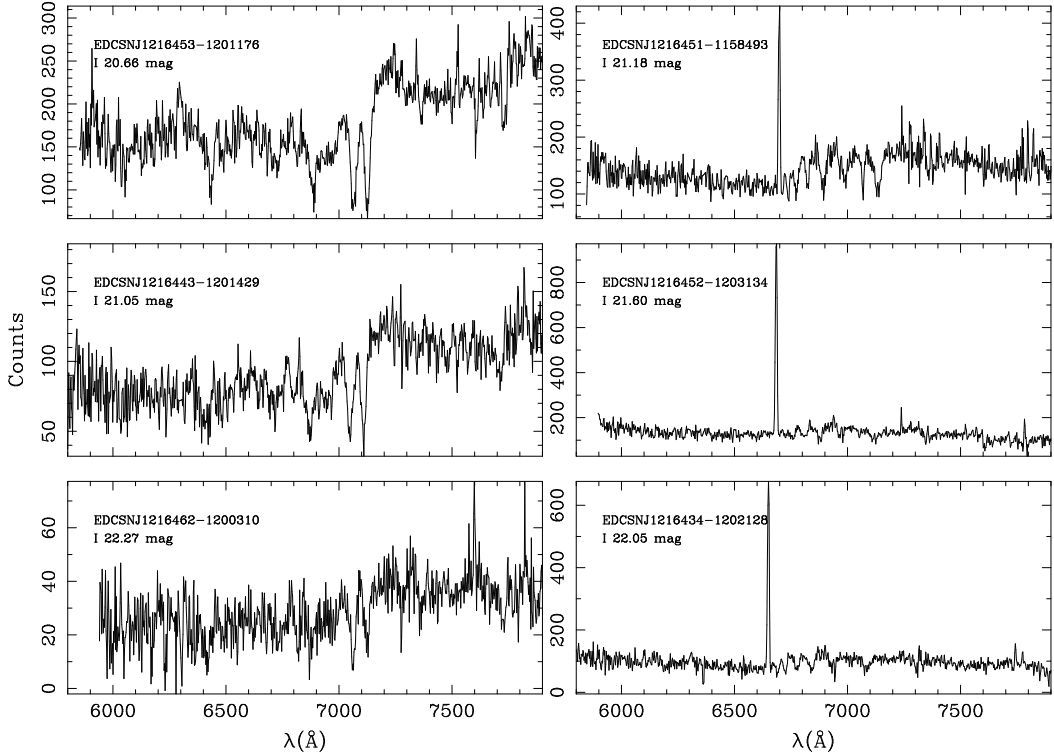}\\
   \caption{ Example spectra (2 pixel binning) of galaxies in the
   cluster Cl~1216 at $z=0.79$. Galaxy IDs and $1^{\prime\prime}$
   aperture I-band magnitudes are given for each spectrum. The bright
   emission line [OII]$\lambda$3727 (clearly visible in the right-hand
   panel figures) was used to determine the galaxy redshift where
   possible; otherwise the redshift was measured using the Calcium H
   and K lines (clearly visible in the interval 7000--7200\AA~of the
   left-hand panel figures). All spectra are presented after
   correction for telluric absorption and before flux
   calibration.\label{spectra1}}

   \end{figure*}

   \begin{figure*}
   \centering
   \includegraphics[width=10.0cm,angle=270]{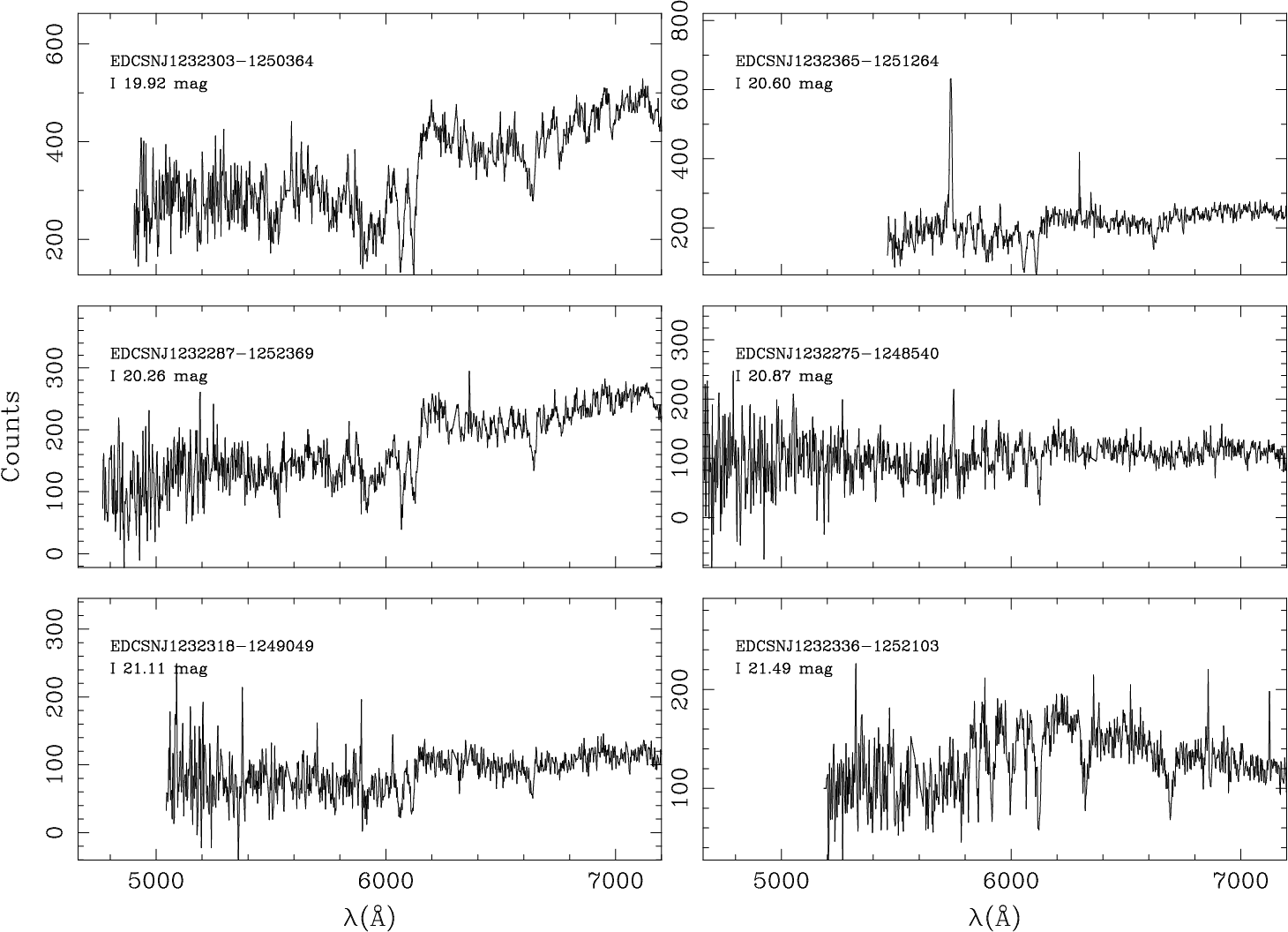}
   \caption{Example spectra (2 pixel binning) of galaxies in the
   cluster Cl~1232 at $z=0.54$. Galaxy IDs and $1^{\prime\prime}$
   aperture I-band magnitudes are given for each spectrum. The bright
   emission line [OII]$\lambda$3727 (clearly visible in the two top
   right-hand panel figures) was used to determine the galaxy redshift
   where possible; otherwise the redshift was measured using the Ca H
   and K lines (clearly visible in the interval 6000--6200\AA~of the
   left-hand panel figures). All spectra are presented after
   correction for telluric absorption and before flux
   calibration. \label{spectra2}}
\end{figure*}
%

\section{Cluster and galaxy redshifts}\label{redshifts}

\subsection{Galaxy redshifts}

We present representative spectra for cluster members of the
high-redshift cluster Cl~1216 (figure \ref{spectra1}) and the
intermediate-redshift cluster Cl~1232 (figure \ref{spectra2}). We show
the spectrum for the brightest cluster galaxy and spectra for an
additional 5 progressively fainter galaxies.

Spectroscopic galaxy redshifts were measured using emission lines
where possible, in particular the [O{\sc ii}]$\lambda$3727 line, or
the most prominent absorption lines, e.g. Calcium K and H lines at
3934$\,${\AA}~and~3968$\,${\AA}. Errors for galaxy redshifts were
calculated using repeat galaxy observations acquired during
spectroscopic runs 1 and 2. Our typical galaxy redshift error was
approximately 0.0003 (90 $\,{\rm km}\,{\rm s}^{-1}$), corresponding to
50 and 60 $\,{\rm km}\,{\rm s}^{-1}$ rest frame at $z = 0.8$ and $z =
0.5$ respectively. We compared our redshift measurements with values
determined using the Fourier Correlation Quotient method of
\citet{Bender:1990} and found a typical intrinsic uncertainty of
0.00011 (18 $\,{\rm km}\,{\rm s}^{-1}$ at $z = 0.8$ rest frame; 22
$\,{\rm km}\,{\rm s}^{-1}$ at $z = 0.5$ rest frame).

Throughout this paper, a galaxy is considered to be a member if its
redshift is within $\pm$(3$\times\sigma_{\rm cluster}$) of the cluster
redshift (cf. expression (\ref{finalzint})), where $\sigma_{\rm
cluster}$ is the cluster velocity dispersion presented in Section
\ref{cvddesc}.

\subsection{The spectroscopic catalogue}\label{speccat}

We present the catalogue of spectroscopic observations of 5 EDisCS
clusters. The tables are available in electronic form at the CDS. The
format of each table is illustrated in Table \ref{spectrocat}. The
tables provide the galaxy ID in column 1; object RA and Dec (J2000) in
columns 2 and 3; I-band magnitude for an aperture of radius
1$^{\prime\prime}$ in column 4; a spectroscopic redshift in column 5;
a membership flag in column 6; and a ``targeting'' flag in column 7.

\begin{table*}
\caption{Example spectroscopic catalogue}
\label{spectrocat}
\begin{tabular}{lccllrr}
\hline
EDisCS object & RA & Dec & I-band & Spectroscopic & Membership & Targeting\\
 & & & magnitude & redshift & flag & flag\\
\hline
EDCSNJ1054316-1147400 & 10:54:31.62 & $-$11:47:40.0 & 21.249 & 0.6908  &  1 & 1 \\
EDCSNJ1054254-1147523:& 10:54:25.38 & $-$11:47:52.3 & 21.380 & 0.6977  &  1 & 1 \\
EDCSNJ1054254-1147523:& 10:54:25.38 & $-$11:47:52.3 & 21.380 & 0.8385  &  0 & 3 \\
EDCSNJ1054260-1148009 & 10:54:26.02 & $-$11:48:00.9 & 21.815 & 0.6802  &  3 & 1 \\
EDCSXJ1054338-1145437 & 10:54:33.81 & $-$11:45:43.7 & 22.90  & 0.6649  &  0 & 1 \\
EDCSNJ1054494-1243321 & 10:54:49.36 & $-$12:43:32.1 & 22.947 & 0.6864::&  0 & 1 \\
EDCSNJ1054504-1243410 & 10:54:50.42 & $-$12:43:41.0 & 21.471 & 0.7309  & 1B & 1 \\
EDCSNJ1054504-1243259 & 10:54:50.44 & $-$12:43:25.9 & 23.875 & 0.225:  &  0 & 3 \\
EDCSNJ1216431-1158113 & 12:16:43.11 & $-$11:58:11.3 & 22.503 & 1.0579: &  0 & 1 \\
EDCSNJ1232389-1252170 & 12:32:38.95 & $-$12:52:17.0 & 21.523 & 0.6809  &  0 & 2 \\
EDCSNJ1232285-1254057 & 12:32:28.54 & $-$12:54:05.7 & 17.873 & 0.0000  &  0 & 4 \\
EDCSNJ1232322-1253003 & 12:32:32.17 & $-$12:53:00.3 & 20.824 & 9.9999  &  0 & 1 \\
\hline
\end{tabular}

\end{table*}

In column 1, a colon (:) indicates that multiple spectra were
extracted for a single photometric object. Two redshifts have an
associated ID with a colon: these two cases are shown in the example
table (Table~\ref{spectrocat}). Our HST ACS image indicates that we
are seeing two galaxies in projection. 3 galaxies were not included in
the final versions of our photometric catalogues due to changes in
object identification criteria. For these galaxies an empirical ID has
been constructed using the galaxy RA and Dec coordinates, and a prefix
of X instead of N; I-band magnitudes are taken from the version of our
photometric catalogues used to prepare the spectroscopic observations.

In column 5, a colon indicates a redshift with a higher than average
uncertainty, while a double colon (::) indicates a doubtful
redshift. $z=0.0000$ corresponds to stars, and $z=9.9999$ is used
where a redshift could not be determined. The uncertainty in the
redshift estimate of the spectra listed with a colon was mostly due to
a lower than usual S/N or sky features affecting the lines most
suitable for the redshift measurements. This resulted in a less
precise identification of the line centres. While the approximate
redshift is confirmed by several lines in these cases, the exact
redshift value cannot be established with the same accuracy as for the
secure redshifts. Those cases with a double colon (6 spectra in total)
are all low S/N spectra, where the redshift estimate had in general to
be based on one single line in emission, whose redshift could not be
confirmed by any other emission or absorption line in the spectrum. In
all of these cases the reality of the observed line in emission was
verified by inspecting the 2D spectra. Only for one spectrum marked by
a double colon was the redshift estimate based on the presence of a
clear break in the continuum intensity, but no specific lines were
identified without doubt due to the low S/N.

The numeric membership flag in column 6 is ``1'' for cluster members
(within 3$\sigma_{\rm cluster}$ of the cluster redshift); ``2'' and
``3'' for galaxies at 3$\sigma_{\rm cluster}$ to 4$\sigma_{\rm
cluster}$ and 4$\sigma_{\rm cluster}$ to 10$\sigma$ from the cluster
redshift, respectively, that can be considered members of a
supercluster region; and ``0'' in all other cases of fore- and
back-ground non-members. The cluster Cl~1054-12 has a secondary
structure 7$\sigma_{\rm cluster}$ away in redshift from the main
structure (see Section 5 and Table~5). In this case a membership flag
is listed both for the main cluster Cl~1054-12A (``1A'', ``2A'',
``3A'', with meaning as above) and for the secondary structure
Cl~1054-12B (``1B'', ``2B'', ``3B''). Where a galaxy can have, for
example, both a ``3A'' and ``1B'' flag, a ``1B'' flag has been
assigned.

Column 7 indicates whether the galaxy was a selected spectroscopic
target, hence was targeted for spectroscopy as a possible member of
the cluster (flag=1); if it was observed but believed to be a
``field'' (non-cluster) galaxy based on the photometric-redshift
criteria (flag=2); if it is a galaxy that happened to fall into the
slit together with a targeted galaxy (flag=3); and if it was targeted
as a star to aid acquisition (flag=4). As described below, this
information can be used to assess the success of our observing
strategy and the effects of the selection criteria adopted.

\subsection{Success rate, completeness and potential selection biases}\label{completeness}

It is important to establish {\it a posteriori} if the criteria used
to select our spectroscopic targets have biased the final
spectroscopic cluster member dataset. As described in Section
\ref{target}, to maximise the number of targeted cluster members and
avoid rejection of possible cluster galaxies, we adopted conservative
criteria based on our photometric redshifts. To recap, our selection
criteria have rejected between 30{\bf \%} and 50\% of the objects in
the full photometric catalogues of our high redshift clusters, and
approximately 35\% of objects in the catalogue of the $\sim 0.5$
cluster (Cl~1232).

\begin{figure*} \centering
   \includegraphics[width=7.5cm]{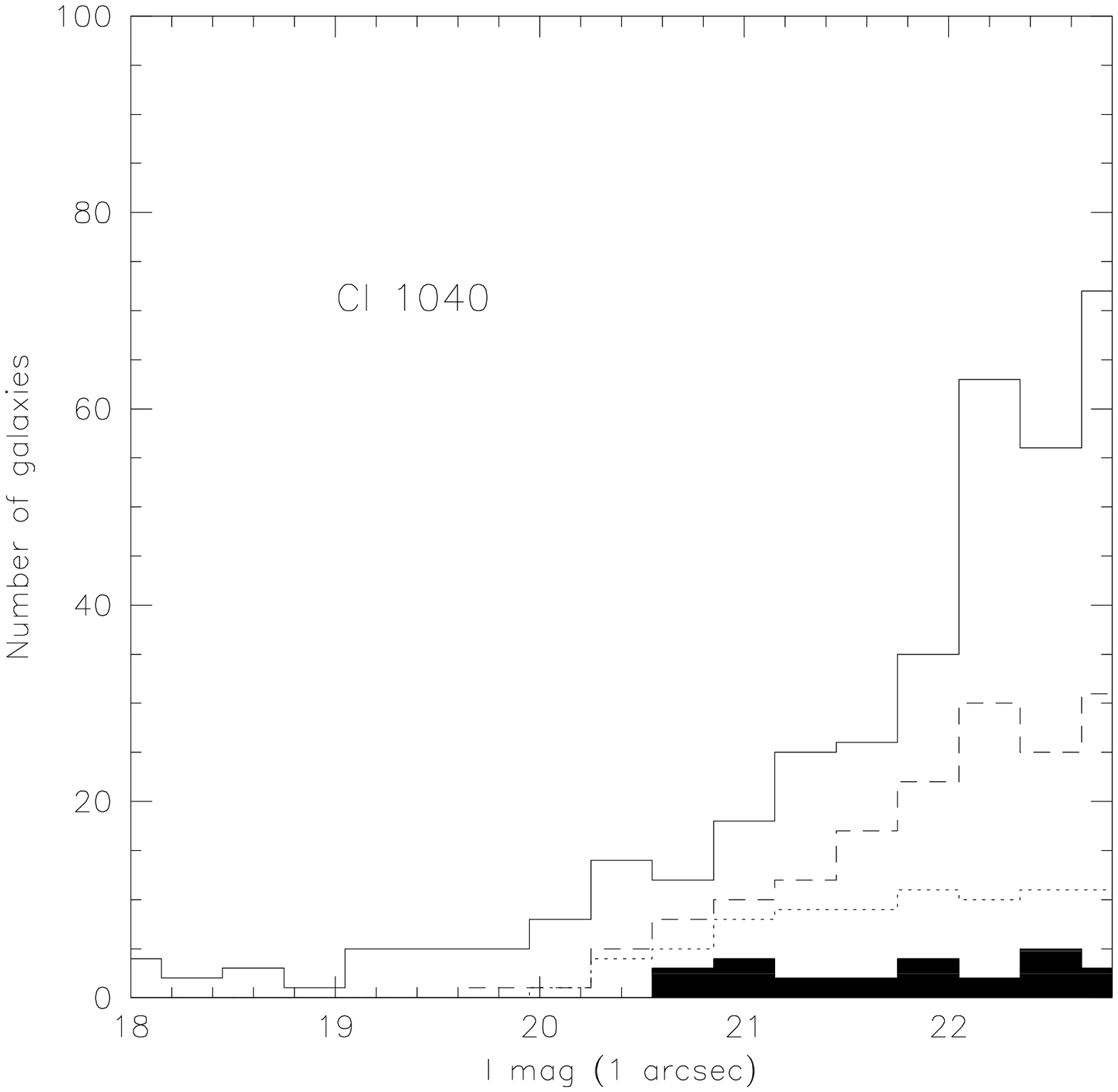}
   \includegraphics[width=7.5cm]{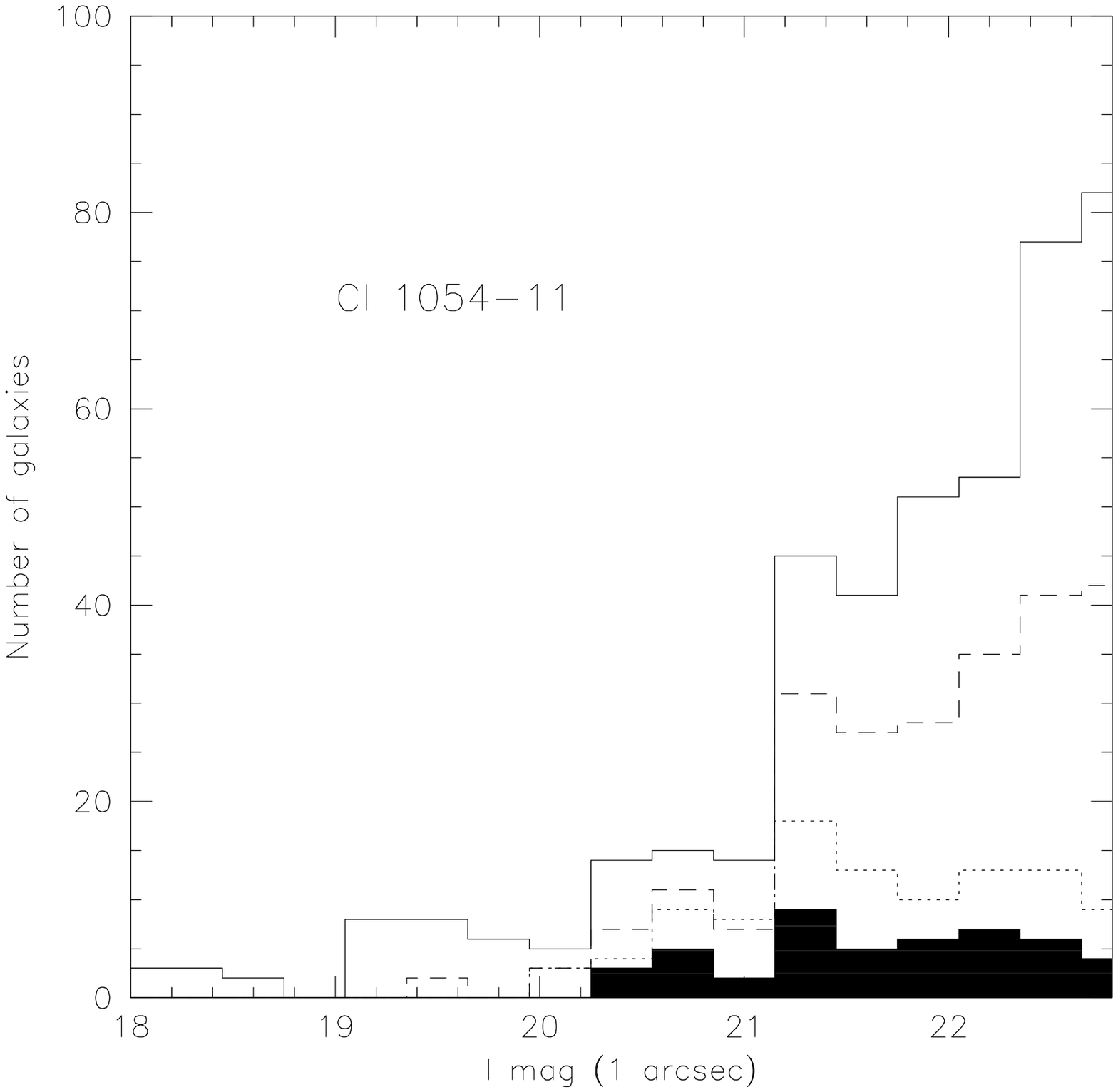}
   \includegraphics[width=7.5cm]{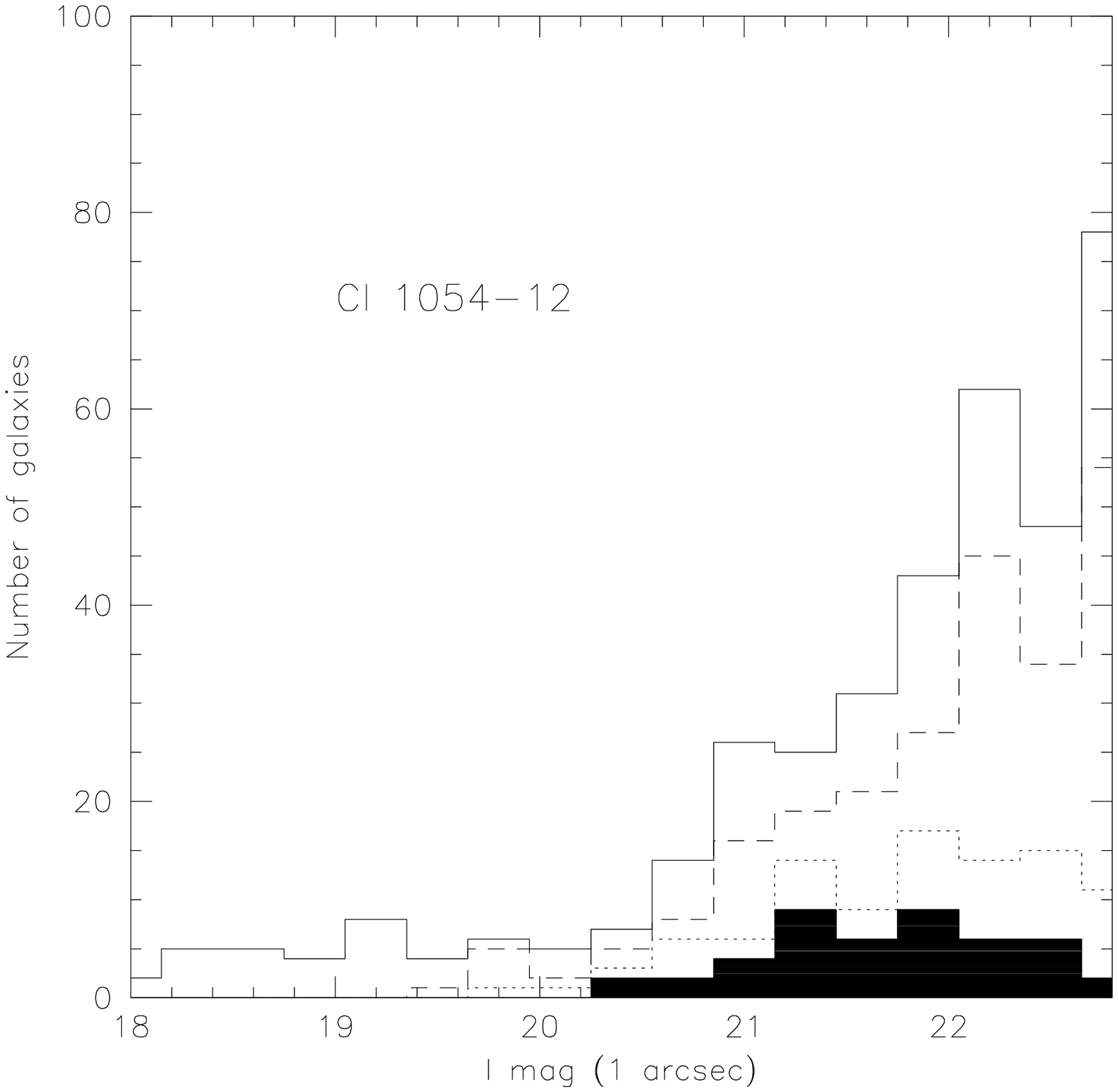}
   \includegraphics[width=7.5cm]{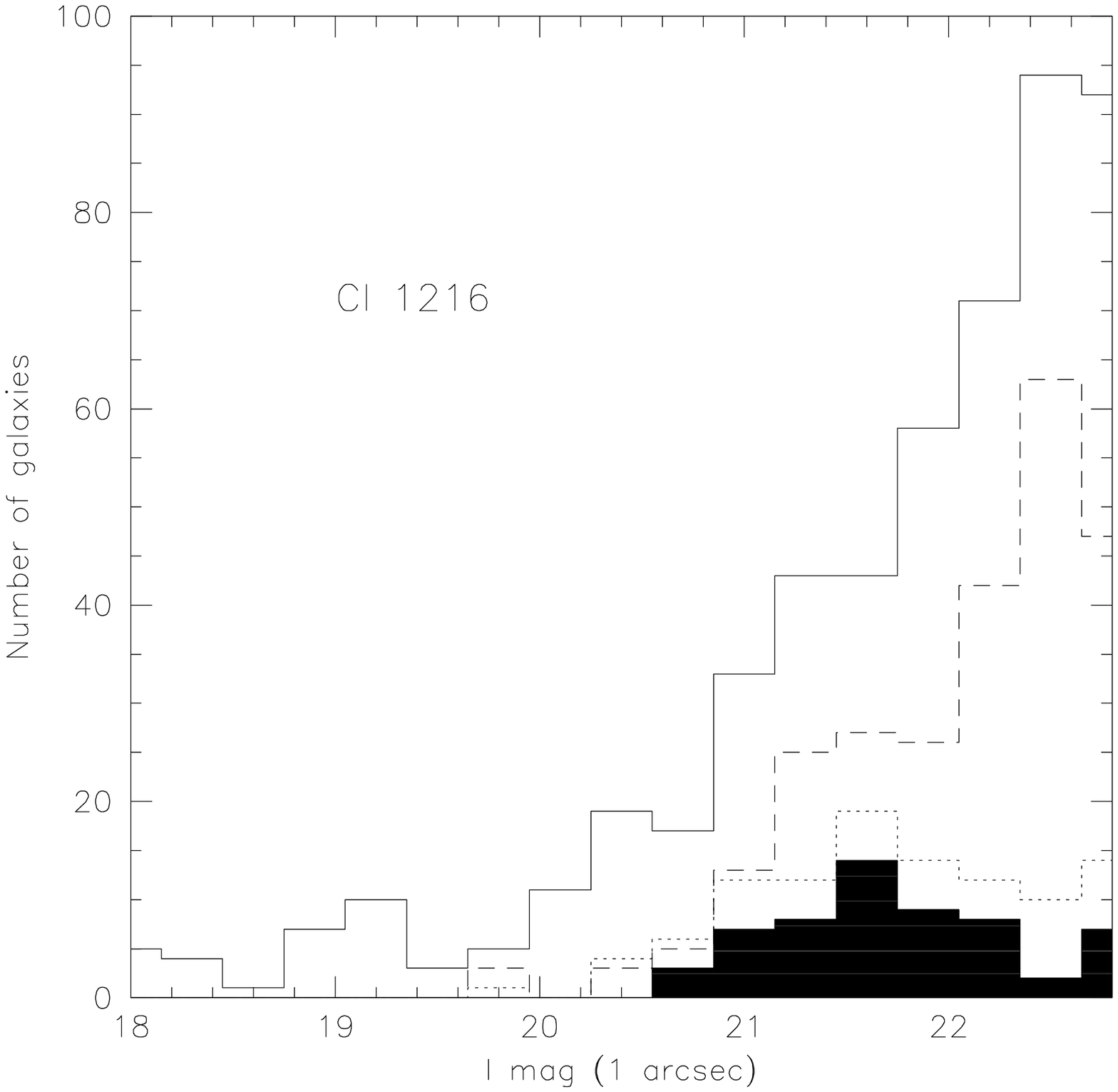}
   \includegraphics[width=7.5cm]{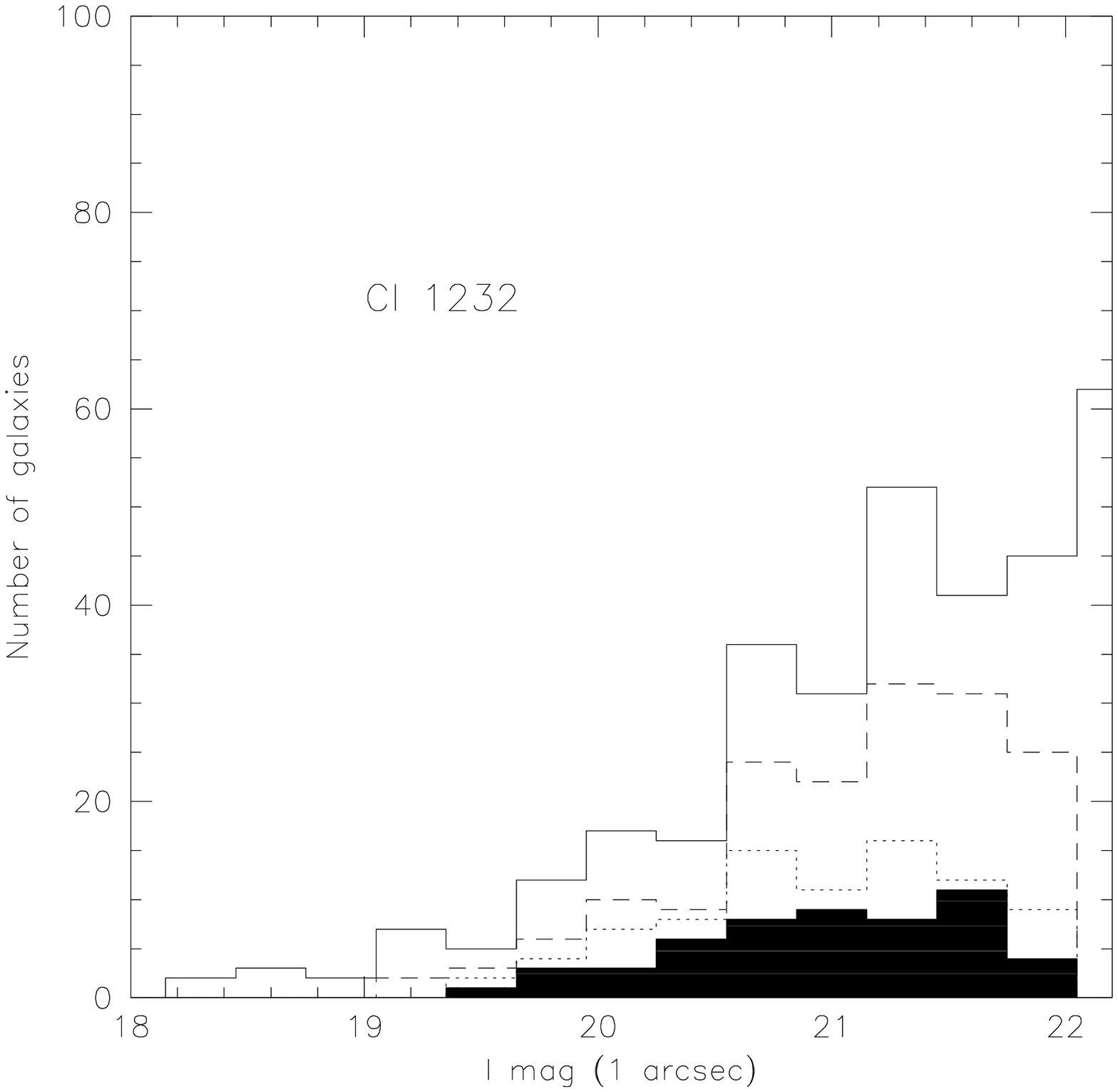} \caption{Magnitude
   distributions for our 5 clusters. The solid, largest histogram
   represents the magnitude distribution of all objects detected by
   SExtractor in the EDisCS fields (the full photometric catalogue in
   the area of sky sampled by the 2002 spectroscopy). The dashed
   histogram is the distribution of objects that were retained as
   potential spectroscopic targets within the area of sky sampled by
   the 2002 spectroscopy after applying the selection criteria
   described in Section \protect\ref{target}. The magnitude
   distribution of selected spectroscopic targets for which a spectrum
   was obtained is shown as a dotted histogram, while that of all
   selected spectroscopic targets confirmed to be cluster members is
   shown as a filled histogram. \label{complete}} \end{figure*}

The solid and dashed histograms in Figure \ref{complete} present the
magnitude distribution of all objects detected by SExtractor in the
EDisCS FORS2 field (i.e. our full photometric catalogues) and of
potential spectroscopic targets decided after applying the selection
criteria described in Section \ref{target}. The same plot illustrates
the magnitude distribution of selected spectroscopic targets for which
a spectrum was acquired (dotted histogram), and of selected
spectroscopic targets which are cluster members according to their
spectroscopic redshift (filled histogram).

Figure \ref{complete} indicates that our membership success rate has
been high. The fraction of objects targeted as cluster members and
then confirmed to be members is approximately 75\% in Cl~1232, 60\% in
Cl~1216, 50\% in Cl~1054-12 and Cl~1054-11, and 30\% in
Cl~1040.\footnote{When computing the success rate, galaxies in
Cl~1054-12A and Cl~1054-12B have been considered together, since the
separation in redshift (0.0195) is far too small to be separated by
any photometric-redshift technique. The lower success rate in Cl~1040
is due to a higher contamination from back- and fore-ground galaxies.} 
The spectroscopic cluster catalogues typically include spectra for
approximately 20\% of all potential spectroscopic targets ($\sim 25\%$
in Cl~1216 and 35\% in Cl~1232). Redshift measurements were obtained
for 91--98\% of objects targeted as cluster members for a given set of
cluster observations. In cases where the spectrum did not yield a
redshift the object was either of very low S/N spectra and among the
faintest of the sample (generally fainter than $I = 22.5$), or
relatively bright objects ($I \sim 20$ or brighter) with a featureless
spectrum that, judging from the shape of the red continuum, most
probably are at a redshift much lower than our clusters.

We recall that due to geometrical constraints in our mask design, 65
slits were assigned to objects {\it rejected} by our cluster member
selection criteria. Importantly {\it none} of these objects was found
to be a cluster member galaxy. Thus our adopted selection criteria
have neither rejected real cluster members nor biased our
spectroscopic catalogue of cluster members relative to a purely I-band
selected sample (Section \ref{target}).

%


\section{Cluster redshifts and velocity dispersions}\label{cvddesc}

In this section we present galaxy redshift histograms (Section
\ref{sect:histograms}) and cluster velocity dispersion measurements
(Section \ref{sect:cvds}) for our 5 EDisCS clusters.

\subsection{Redshift histograms}\label{sect:histograms}

\begin{figure*} \centering
   \includegraphics[width=7.5cm]{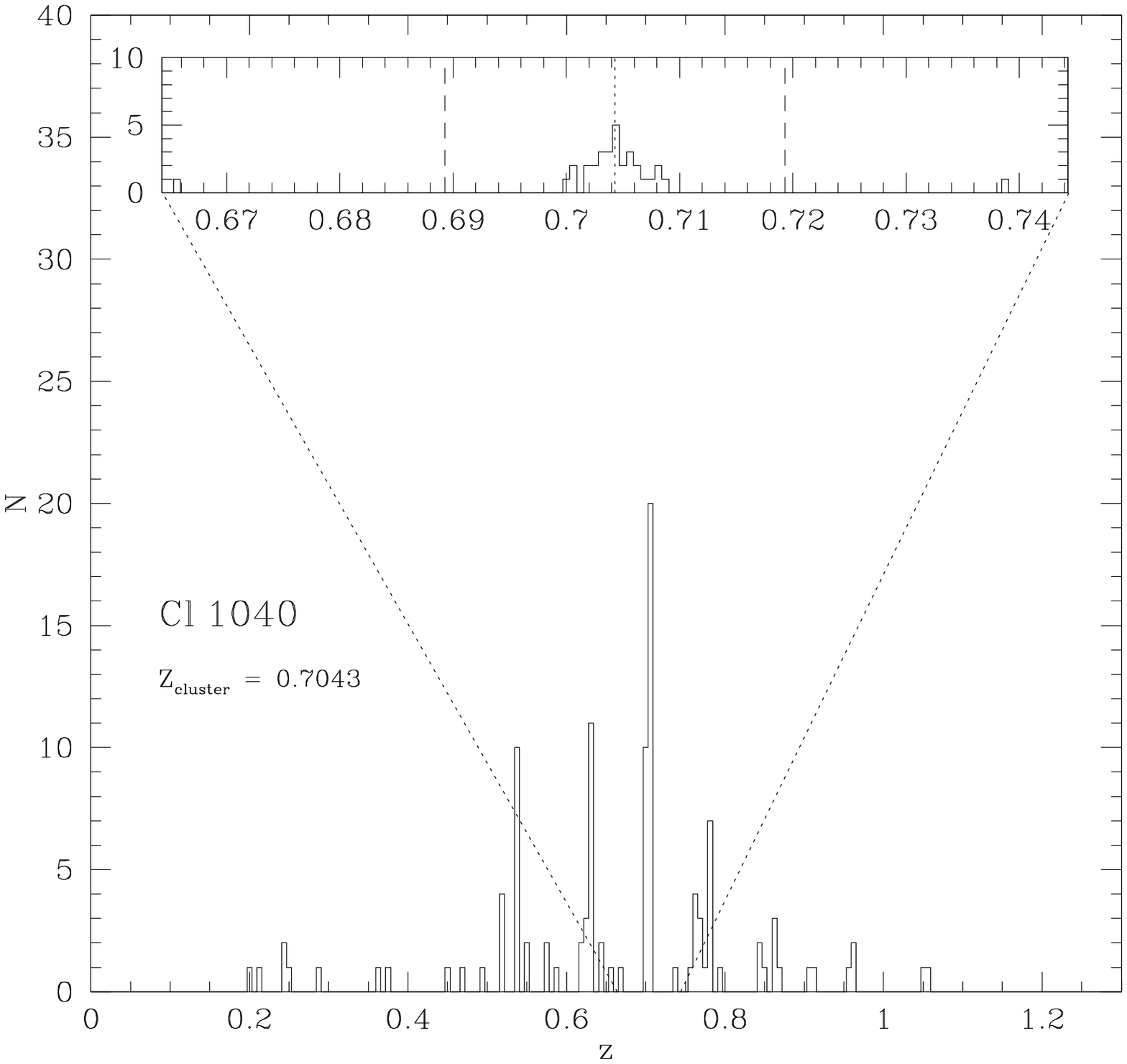}
   \includegraphics[width=7.5cm]{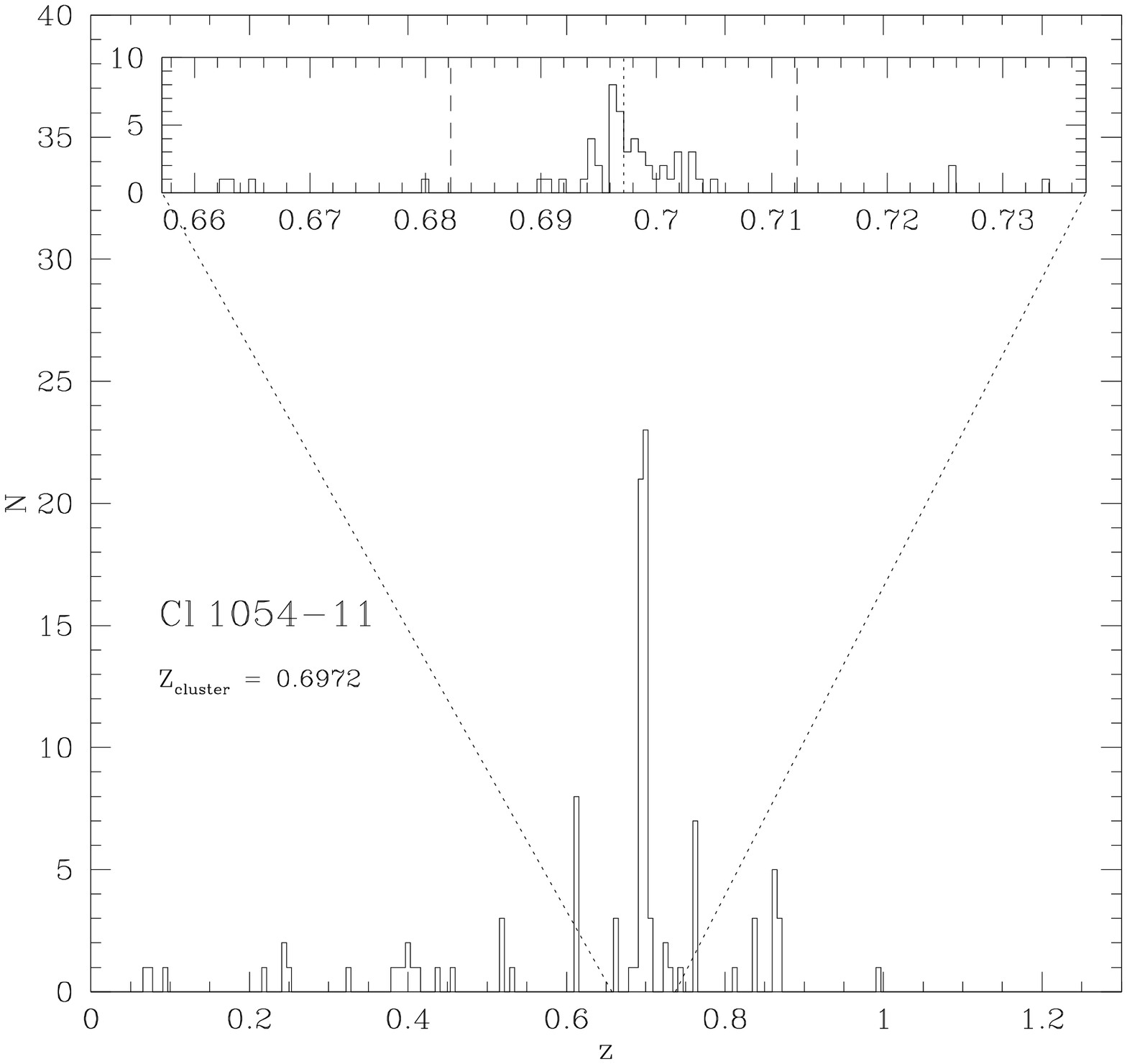}
   \includegraphics[width=7.5cm]{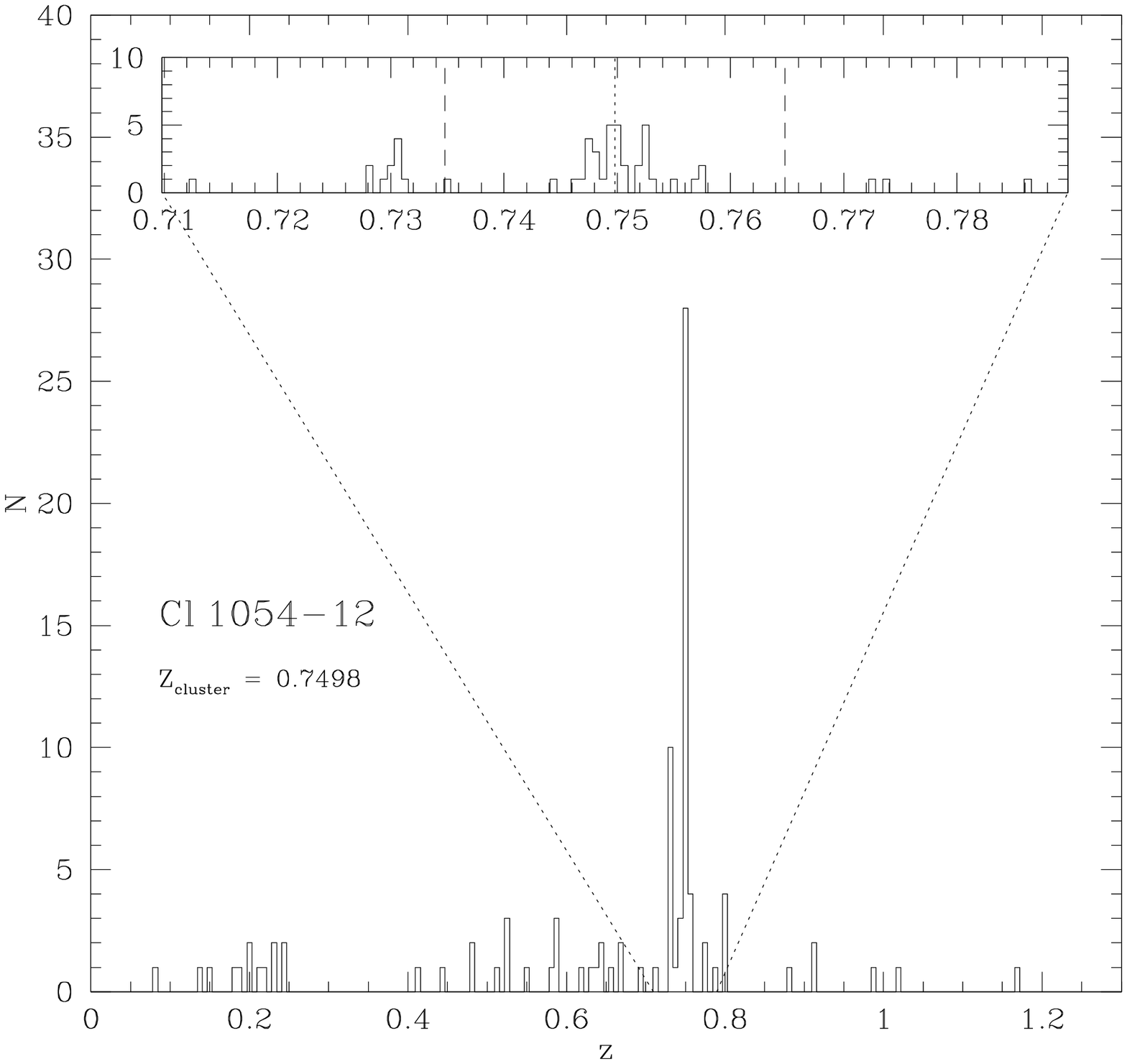}
   \includegraphics[width=7.5cm]{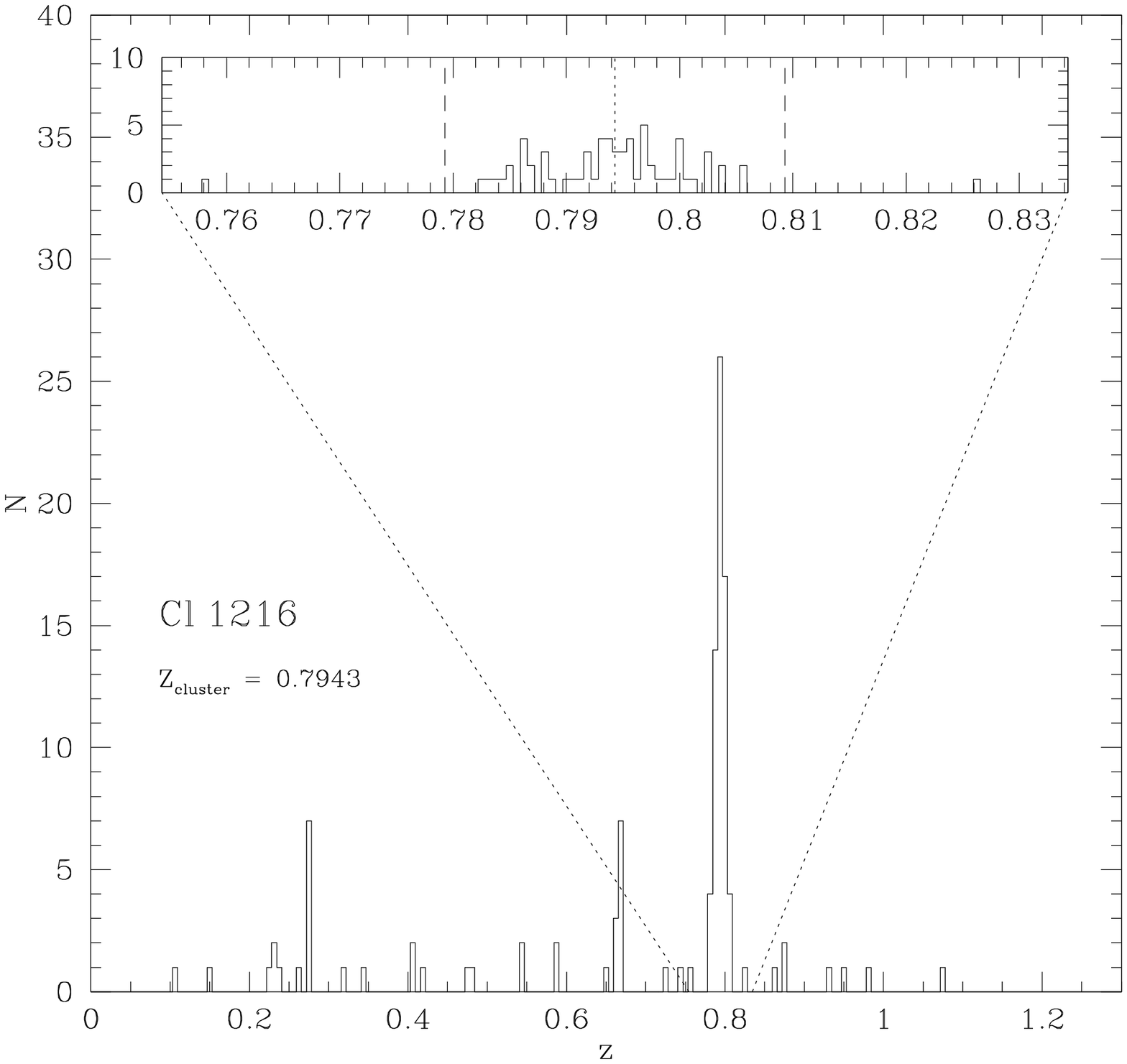}
   \includegraphics[width=7.5cm]{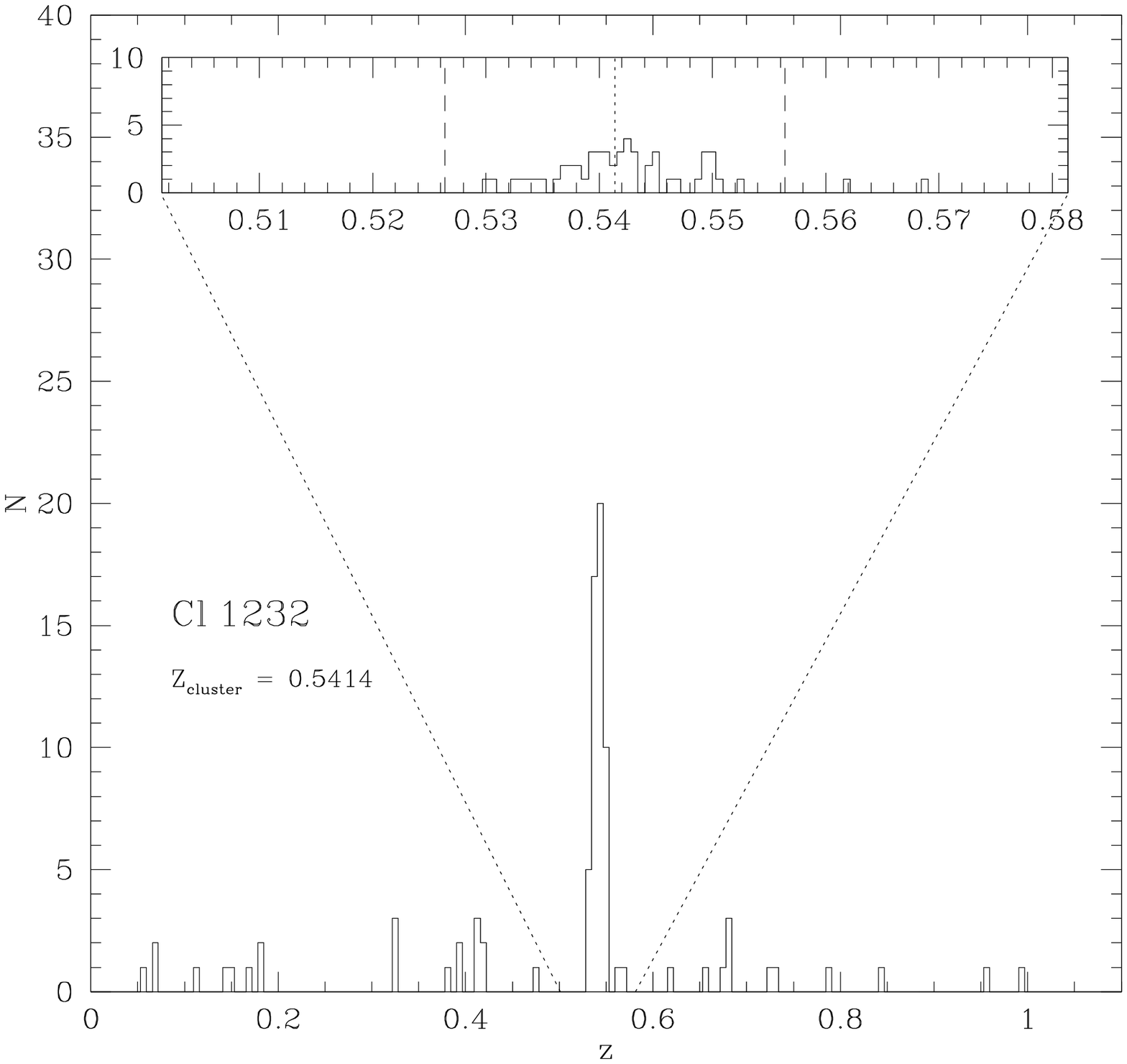}

    \caption{\label{histograms} Redshift histograms based on the long
    spectroscopic exposures of our 5 EDisCS clusters. An inset panel
    in each plot shows a zoomed histogram for the redshift interval
    \protect$z_{\rm cluster} - 0.04 \leq z \leq z_{\rm cluster} +
    0.04$, where the cluster redshift \protect$z_{\rm cluster}$ is
    indicated by a dotted line. Dashed lines indicate the interval
    \protect$z_{\rm cluster} - 0.015 \leq z \leq z_{\rm cluster} +
    0.015$. Redshift bins are 0.00625 (1874 $\,{\rm km}\,{\rm s}^{-1}$
    observed frame) for the main plot and 0.000625 for the inset
    plot. In all cases the redshift distribution is dominated by one
    peak at the cluster redshift.}
\end{figure*}

In Figure \ref{histograms} we show redshift histograms obtained from
the long spectroscopic observations of each cluster. The redshift
distribution is clearly dominated by one main peak in redshift space
at the cluster redshift. For Cl1054-12, a secondary structure is
evident very close to the main cluster peak. For Cl1232 there appears
to be substructure in its velocity histogram. It is not immediately
obvious that these structures are spatially offset from each other on
the sky (see Section \ref{substructure}).

\subsection{Cluster velocity dispersions}\label{sect:cvds}

We present the measurement of cluster velocity dispersion
($\sigma_{\rm cluster}$) for each cluster.

A widely adopted estimator for cluster velocity dispersions is the
biweight statistic discussed by \citet{Beers_etal:1990} (hereinafter
BFG90). For a typical dataset of 20 to 50 galaxies BFG90 show that
the {\it biweight scale estimator} is a robust estimator of
distribution spread and demonstrate its insensitivity to outlying
points.

For each cluster, we estimate $\sigma_{\rm cluster}$ using biweight
statistics: $\sigma_{\rm cluster}$ is taken to be the biweight
estimator of scale for an input redshift list and its asymmetric
errorbars are estimated from Monte Carlo bootstrap simulations
(BFG90). Spectroscopic redshift measurements of uncertain quality
(with a colon or double colon in Table~4) were rejected when
estimating these quantities.

An initial estimate of mean cluster redshift ($z_{\rm init}$) was made
using redshifts judged to belong to the main peak of each cluster
redshift histogram.

Redshifts $z_{i}$ within an interval of:

\begin{equation}
(z_{\rm init} - 0.015) \leq z_{i} \leq (z_{\rm init} + 0.015)
\end{equation}

\noindent were used to obtain a first estimate of
$\sigma_{\rm cluster}$. To transform to the cluster rest-frame,
redshifts were multiplied by the speed of light, ${\rm c}$, to create
recessional velocities $v_{{\rm init}_{i}}$; the median of all these
recessional velocities ($v_{\rm median}$) was then found, and
velocities in the cluster reference frame were taken to be:

\begin{equation}\label{rest}
v_{rest_{i}} = (v_{init_{i}} - v_{\rm median})/(1 + \frac{v_{\rm median}}{c}).
\end{equation}

\noindent An improved estimate of $\sigma_{\rm cluster}$ was
taken to be the biweight estimator of scale for all these $v_{{\rm
rest}_{i}}$. Its uncertainty was estimated as the 1-$\sigma$ bootstrap
confidence interval. Next a 3-$\sigma$ clipping was performed i.e. all
input $v_{{\rm rest}_{i}}$ outside the interval:

\begin{equation}\label{finalzint}
-(3\times\sigma_{\rm cluster}) \leq v_{rest_{i}} \leq
(3\times\sigma_{\rm cluster})
\end{equation}

\noindent were rejected and the measurement of $\sigma_{\rm cluster}$
was repeated on the remaining velocity set. This step was iterated
until expression (\ref{finalzint}) held true for all input $v_{{\rm
rest}_{i}}$. The uncertainty of the final $\sigma_{\rm cluster}$
measurement was again estimated as the 1-$\sigma$ bootstrap confidence
interval. We define {\it cluster members} to be those galaxies for
which expression (\ref{finalzint}) holds after the final iteration.

In Table \ref{cvdtable} we give the results of this exercise. The {\it
cluster redshift} given here, $z_{\rm cluster}$, is the median
redshift of all {\it cluster members}.

\begin{table}
\begin{center}
\begin{tabular}{lcccc} 
\hline
    Cluster name & $z_{\rm cluster}$ & $\sigma_{\rm cluster} \pm \delta\sigma_{\rm cluster}$ & No. of & No. of 
     \\
& & & members I & members II\\
     (1)   &  (2)  &  (3)  &  (4) &  (5) \\ \hline
 Cl\,1040     & 0.7043  &  418 $_{-46}^{+55}$  & 30 & 30 \\
 Cl\,1054-11  & 0.6972  &  589 $_{-70}^{+78}$  & 48 & 49 \\
 Cl\,1054-12  & 0.7498  &  504 $_{-65}^{+113}$ & 35 & 36 \\
 Cl\,1216     & 0.7943  & 1018 $_{-77}^{+73}$  & 66 & 67 \\
 Cl\,1232     & 0.5414  & 1080 $_{-89}^{+119}$ & 52 & 54 \\
\end{tabular}
\medskip
\end{center}
\caption{
Cluster member numbers, redshifts and velocity dispersions for our 5
EDisCS clusters. The cluster name is indicated in column 1. In column
2 we provide the cluster redshift. In column 3, cluster velocity
dispersions ($\sigma_{\rm cluster}$) are shown together with their
errors in $\,{\rm km}\,{\rm s}^{-1}$. In column 4 we indicate the
number of cluster members considered in the measurement of cluster
velocity dispersion (see Section \protect\ref{sect:cvds}). In column 5
the number of cluster members is revised to include members with
redshifts of higher than average uncertainty and doubtful redshifts
(see Sections
\protect\ref{speccat} and \protect\ref{sect:cvds}).}
\label{cvdtable}
\end{table}


\section{Cluster substructure}\label{substructure}

\begin{figure*}

   \centering
   \includegraphics[width=6cm]{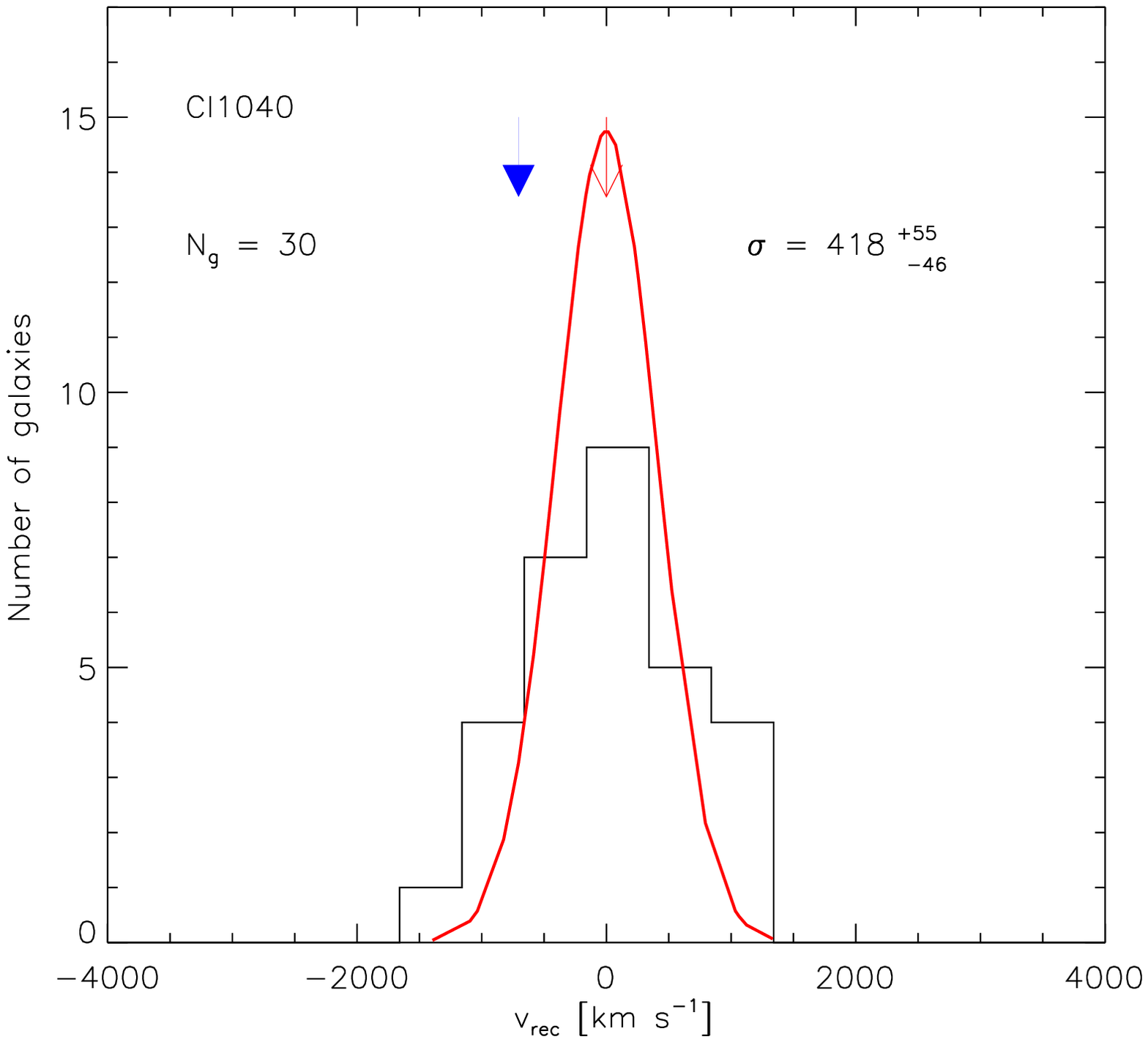}
   \includegraphics[width=6cm]{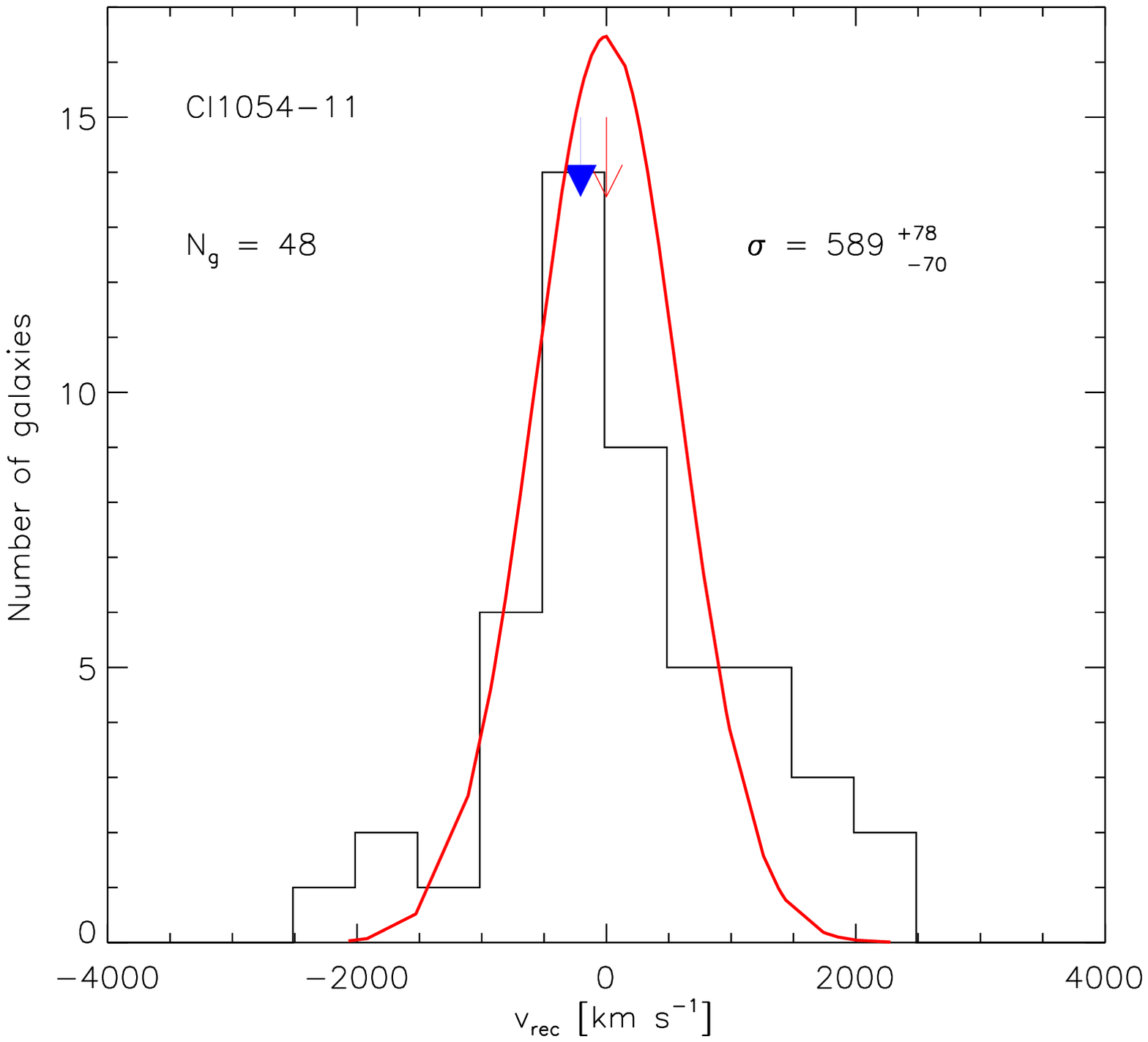}
   \includegraphics[width=6cm]{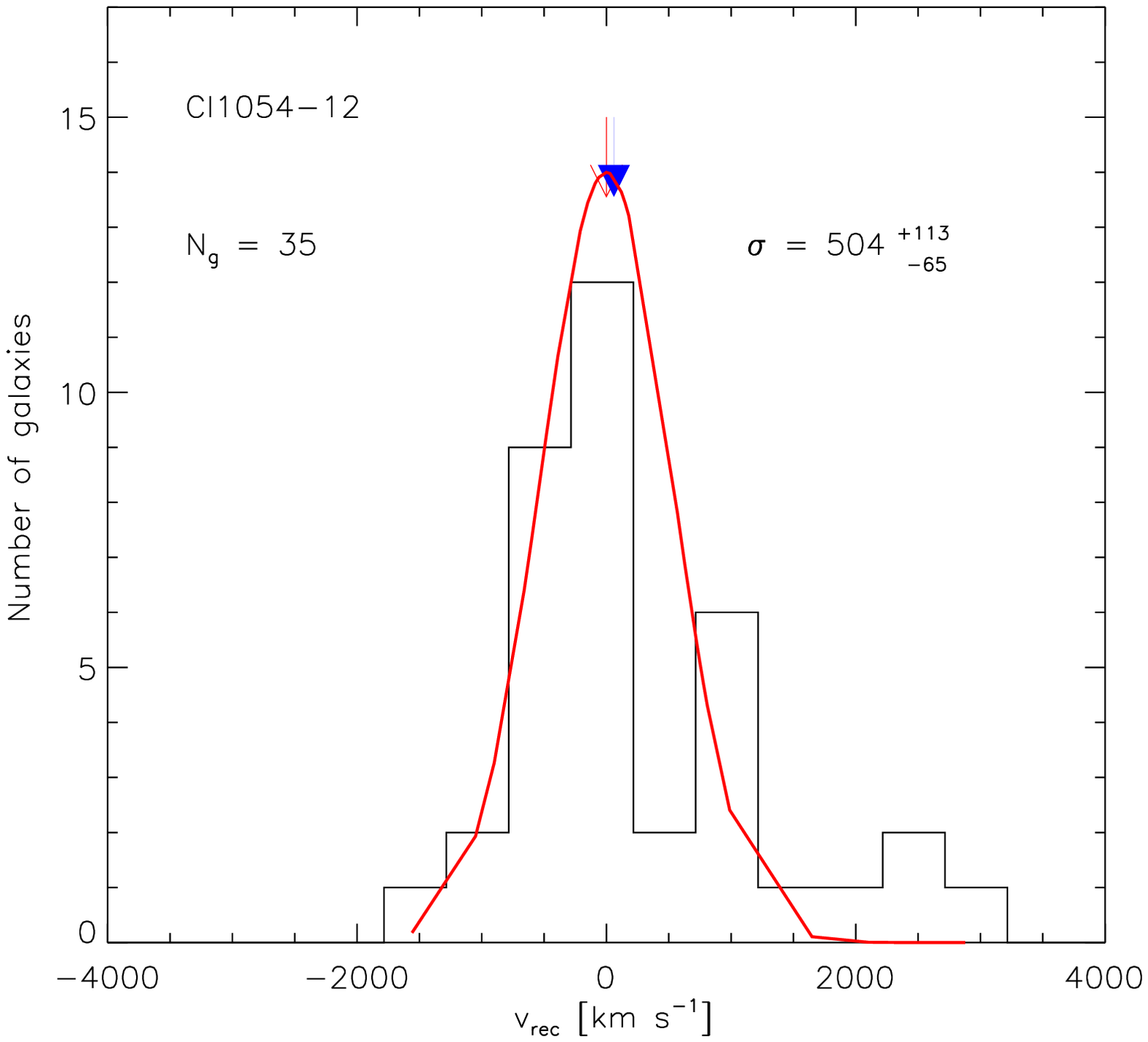}
   \includegraphics[width=6cm]{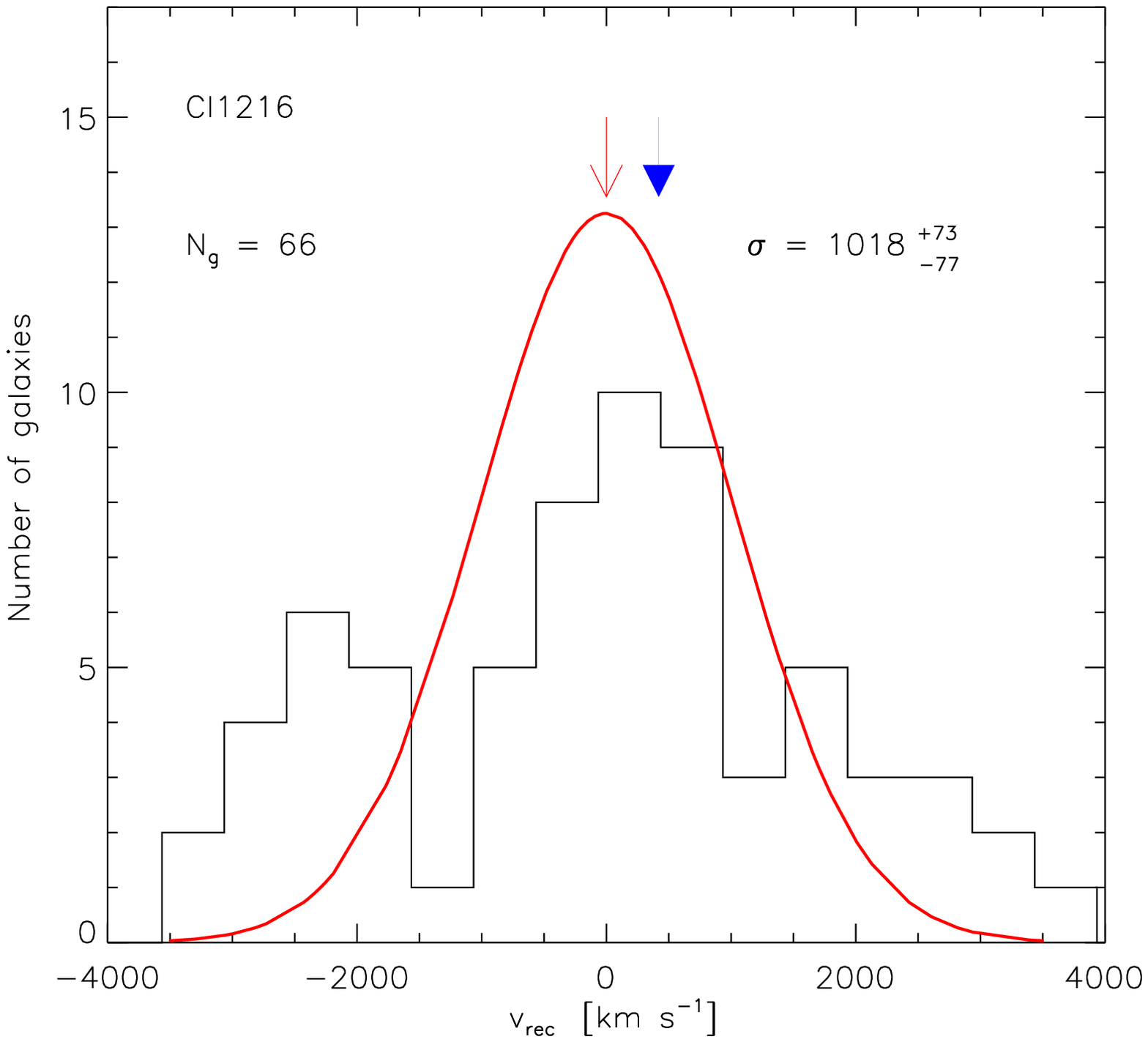}
   \includegraphics[width=6cm]{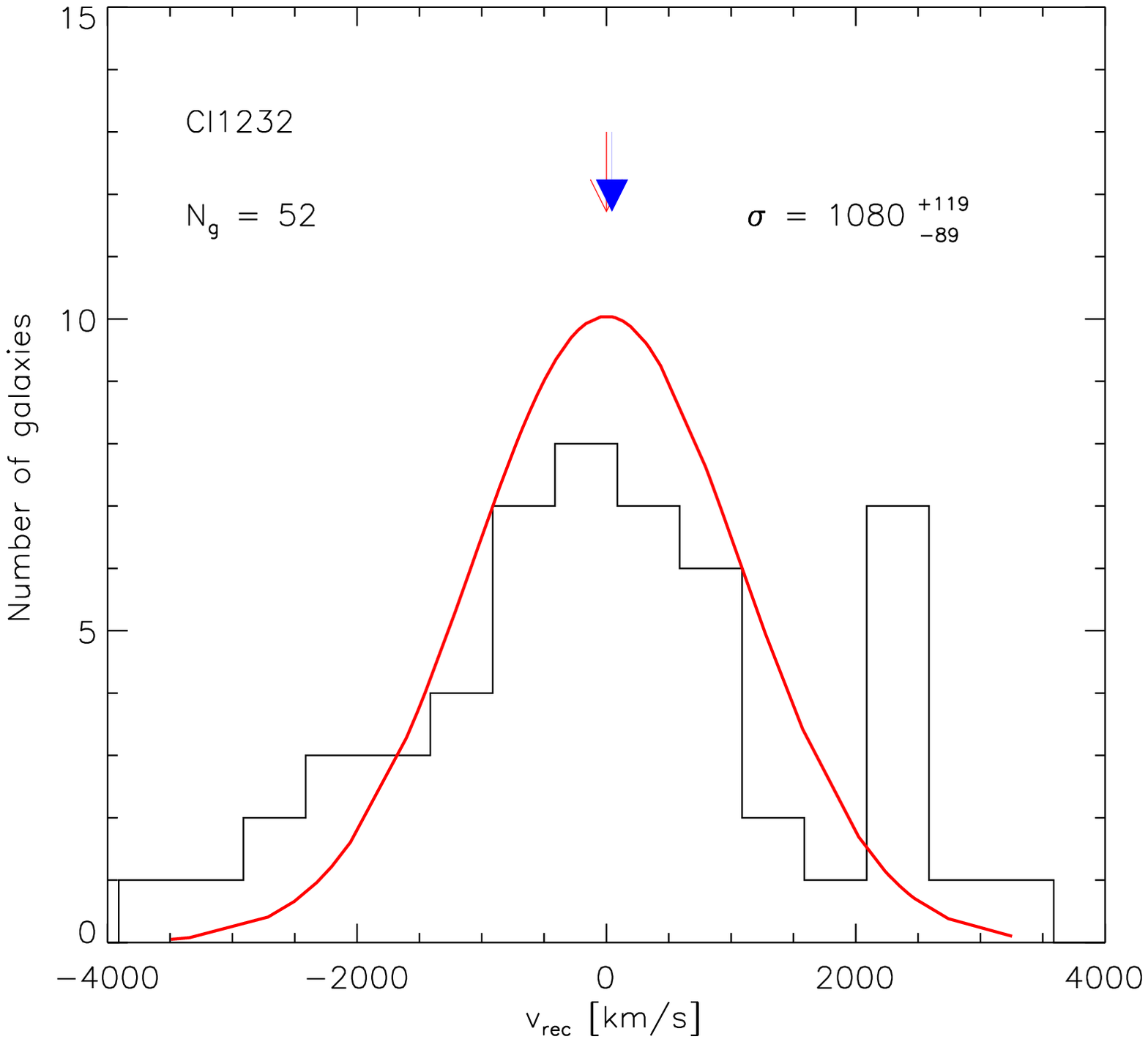}

\caption{Velocity histograms for the five clusters discussed in this paper. 
  The red open symbol arrows mark the median recessional velocity of
  all the spectroscopically-confirmed cluster members, while the blue
  filled arrows mark the recessional velocity of the BCGs. The red
  Gaussians correspond to the measured velocity dispersion.}
\label{fig:velhistohigh}

\end{figure*}

Figure \ref{fig:velhistohigh} presents zoomed-in velocity histograms
for our five clusters using 500 $\,{\rm km}\,{\rm s}^{-1}$ bins. In
each panel, the red arrow marks the median recessional velocity of the
cluster members (i.e. the cluster redshift defined above), and the
blue arrow marks the recessional velocity of the BCG. A Gaussian
centred at the median recession velocity and of width corresponding to
the cluster velocity dispersion is plotted as reference. The velocity
histograms show different levels of deviation from a Gaussian,
providing evidence of complex dynamical structure and suggesting that
some clusters are far from equilibrium. In this Section, we assess
substructure within each cluster first based on XY position maps of
cluster members for different redshift intervals (Section
\ref{desxyplots}) and then using the Dressler-Shectman test (Section
\ref{DSanalysis}).

\subsection{XY position diagrams }\label{desxyplots}

\begin{figure*}
   \centering
   \includegraphics[width=7.5cm]{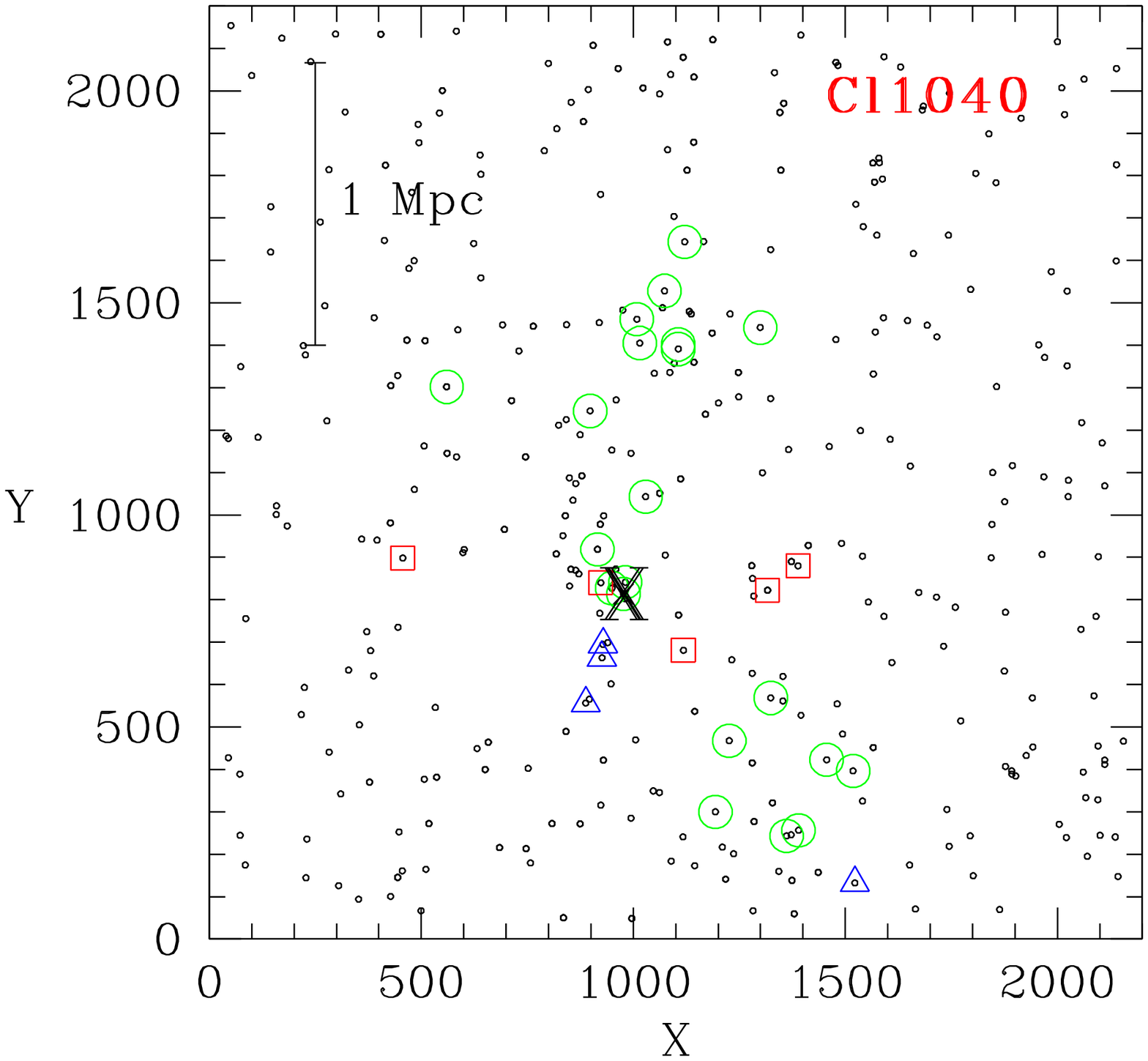}
   \includegraphics[width=7.5cm]{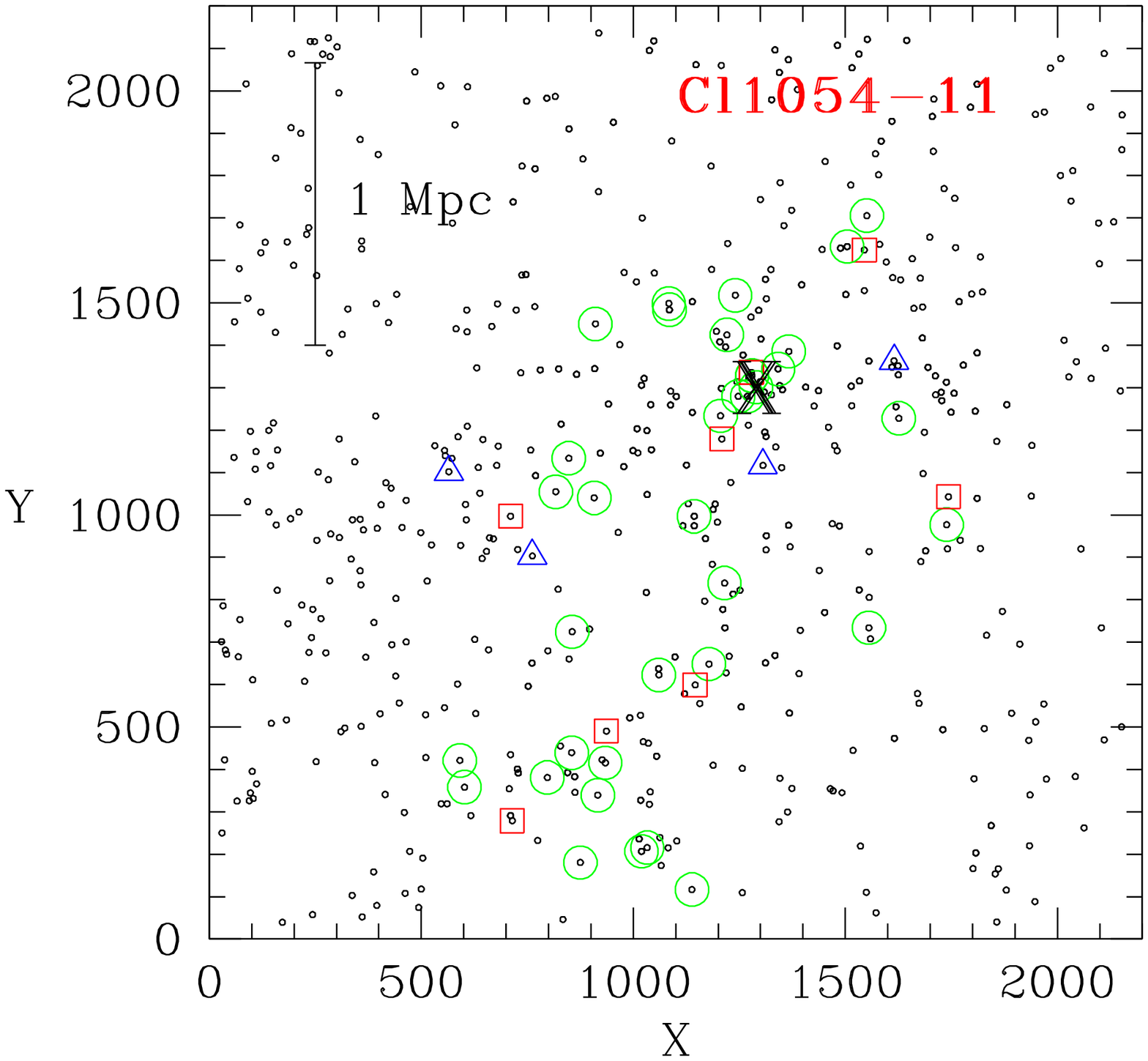}
   \includegraphics[width=7.5cm]{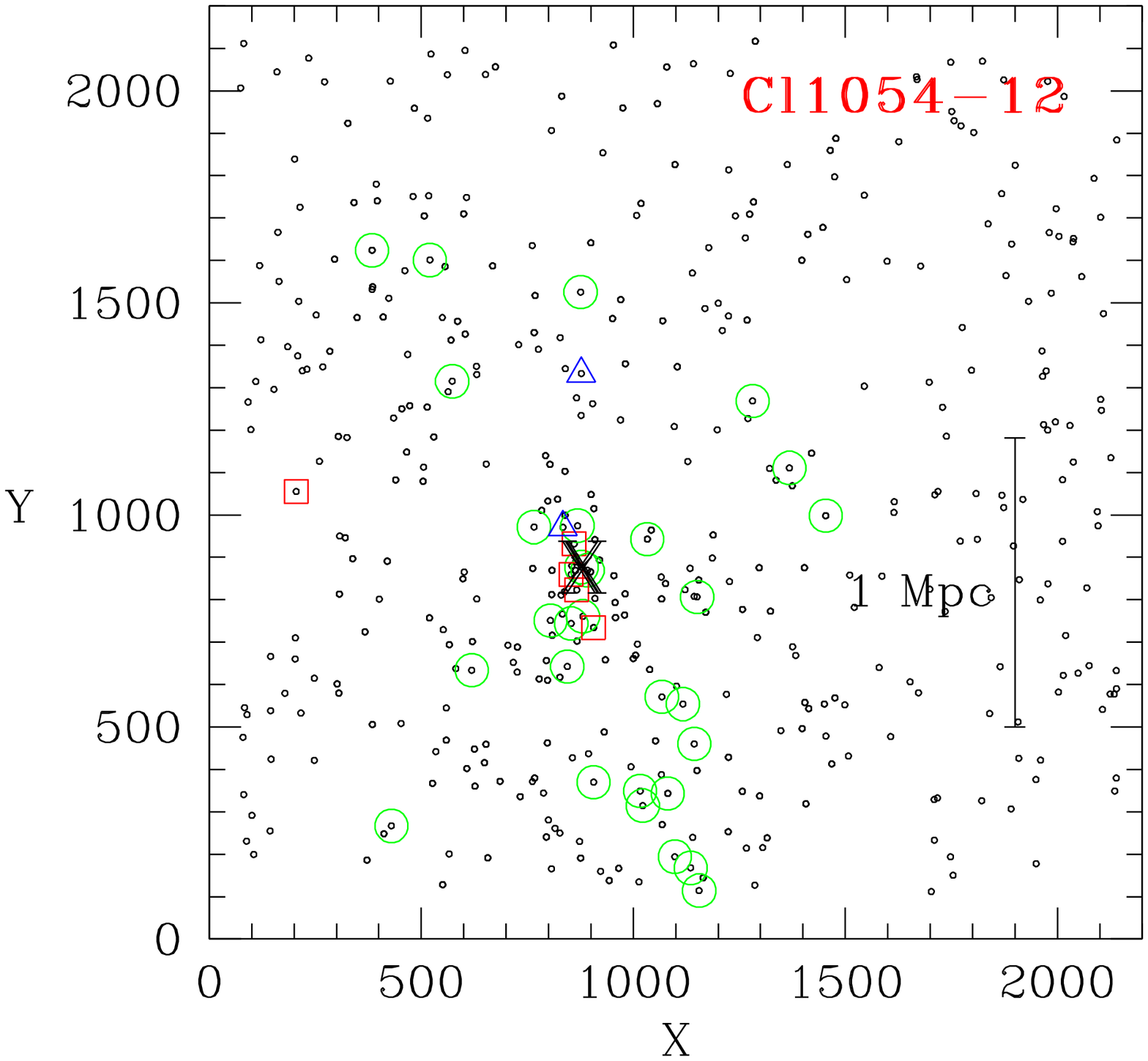}
   \includegraphics[width=7.5cm]{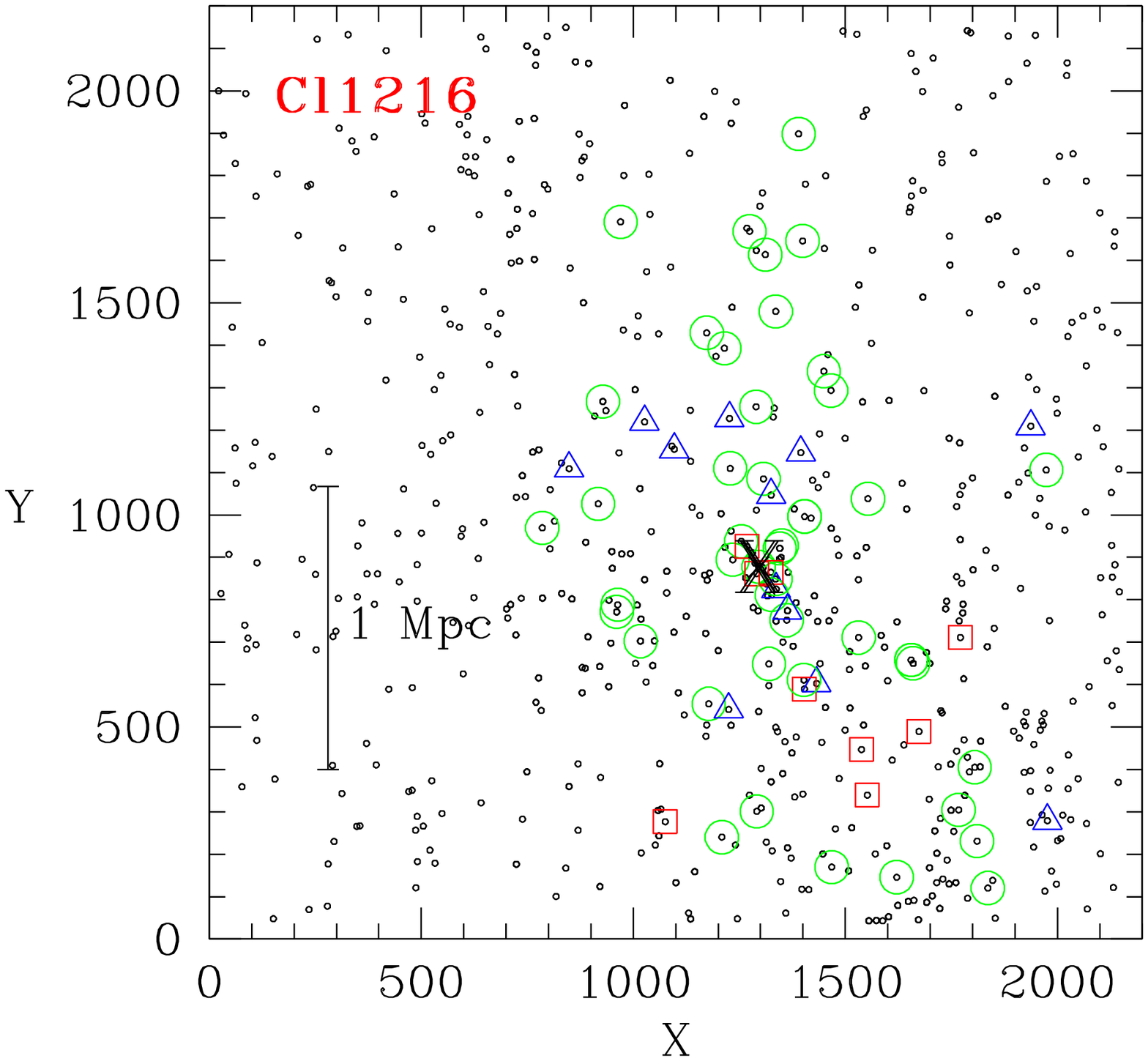}
   \includegraphics[width=7.5cm]{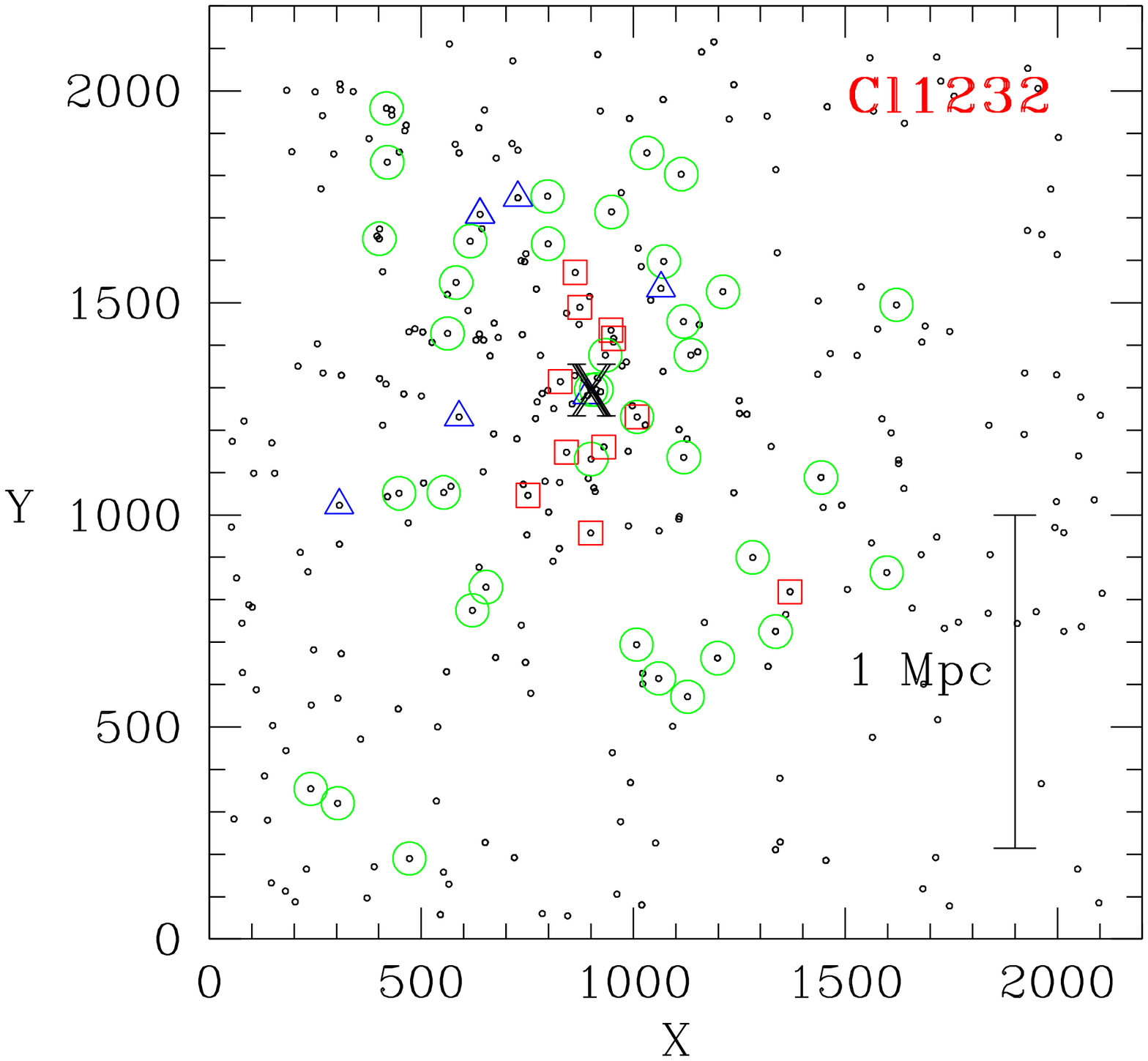}

   \caption{Plots of galaxy positions. Axis scales are in units of CCD
   pixels (0.2$^{\prime\prime}$/pixel). The cluster name is indicated
   at the top of each plot. Large open symbols show the positions of
   cluster members: green circles indicate galaxies within $\pm$ one
   $z_{\sigma}$ of the cluster redshift, red squares indicate galaxies
   whose redshift is more than one $z_{\sigma}$ higher than z$_{\rm
   cluster}$, while blue triangles indicate galaxies whose redshift is
   lower than z$_{\rm cluster}$ by more than one $z_{\sigma}$ where
   $z_{\sigma}$ is the redshift interval corresponding to the cluster
   velocity dispersion in observed frame. Small circles are galaxies
   without spectroscopy selected from our I-band photometric
   catalogues to be possible cluster members. The position of the
   brightest cluster galaxy (BCG) is indicated by a black cross
   symbol. A physical distance of 1 Mpc (assuming $\Omega$ = 0.3,
   $\Lambda$ = 0.7 and $h$ = 0.7) is shown in each plot.}
   \label{fig:xyplots} \end{figure*}

In Figure \ref{fig:xyplots} we present XY position maps for galaxies
within the I-band magnitude range for which our spectroscopy was
completed and we consider as possible cluster members on the basis of
their photo-$z$'s. Objects identified as non-members from our
spectroscopy are excluded. For each cluster the position of the BCG is
indicated by both a black cross and a colour symbol. We study cluster
structure in three dimensions by subdividing cluster members into
three groups corresponding to different redshift intervals. Members
indicated by green circles are all galaxies with redshift $z_{i}$ in
the interval

\begin{displaymath}
(z_{\rm cluster} - z_{\sigma}) \le z_{i} \le (z_{\rm cluster} + z_{\sigma})
\end{displaymath}

\noindent where $z_{\sigma} = \frac{\sigma_{\rm cluster}}{c}\times(1 +
z)$ is the redshift interval corresponding to the cluster velocity
dispersion $\sigma_{\rm cluster}$ presented for each cluster in
Section \ref{sect:cvds}.

Similarly blue triangles indicate cluster members with redshift
$z_{i}$ such that

\begin{displaymath}
z_{i} < (z_{\rm cluster} - z_\sigma)
\end{displaymath}
and red squares, cluster members with redshift $z_{i}$ such that
\begin{displaymath}
z_{i} > (z_{\rm cluster} + z_\sigma).
\end{displaymath}

For Cl~1054-12, some clustering of galaxies indicated by red and green
symbols may be present close to the cluster centre. For Cl~1216,
there are two clear substructures, one given by galaxies with red and
green symbols at cluster centre and another less concentrated
structure offset towards low X and high Y (northeast of the BCG),
shown mainly by blue symbols. For Cl~1232 an extended substructure
indicated mainly by red symbols is clear towards the cluster
centre. No clear spatial segregation is apparent for any redshift
intervals in Cl~1054-11 and Cl~1040.

\subsection{The Dressler-Shectman test}\label{DSanalysis}

   \begin{figure*}
   \centering
   \includegraphics[width=7.5cm]{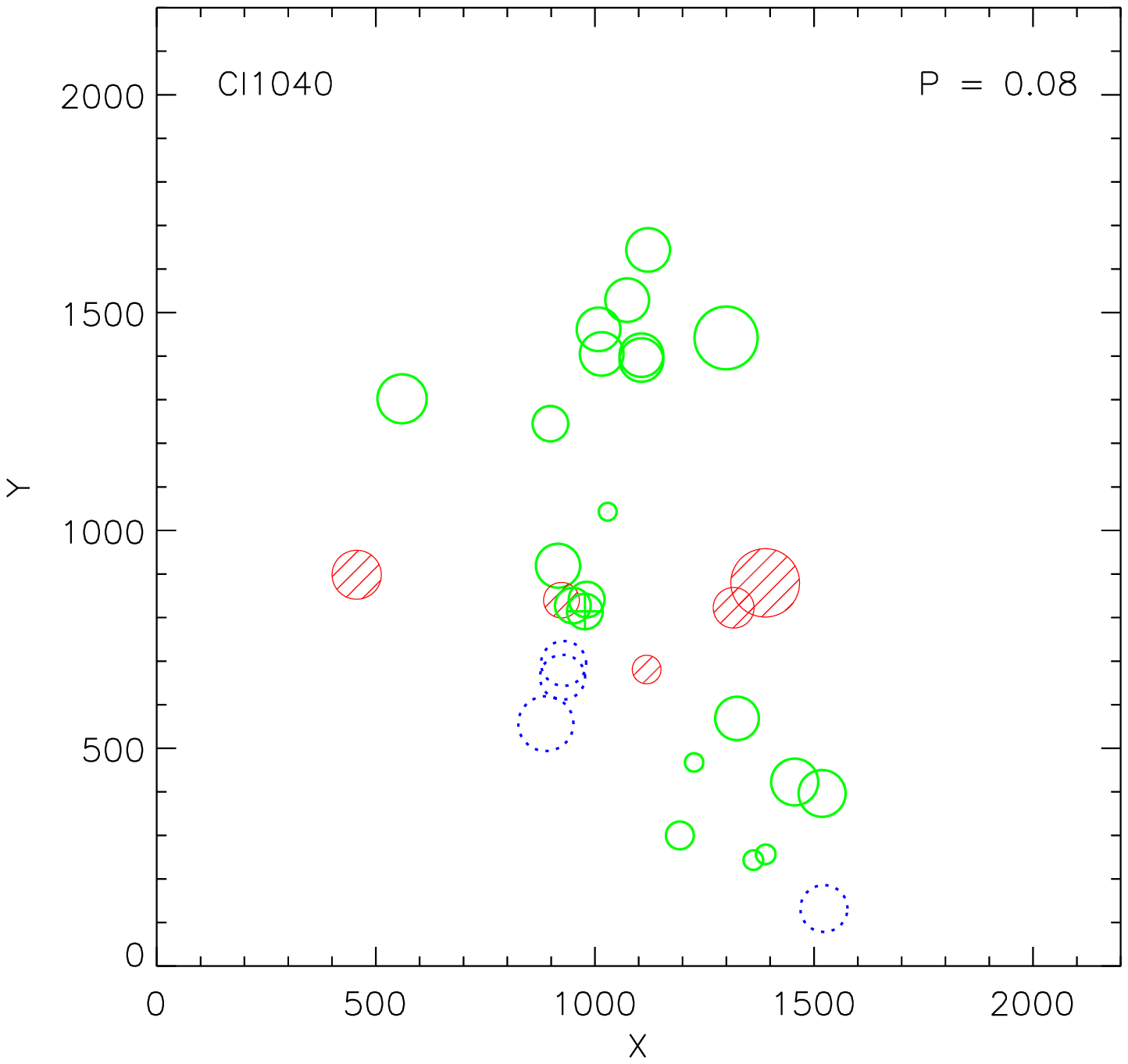}
   \includegraphics[width=7.5cm]{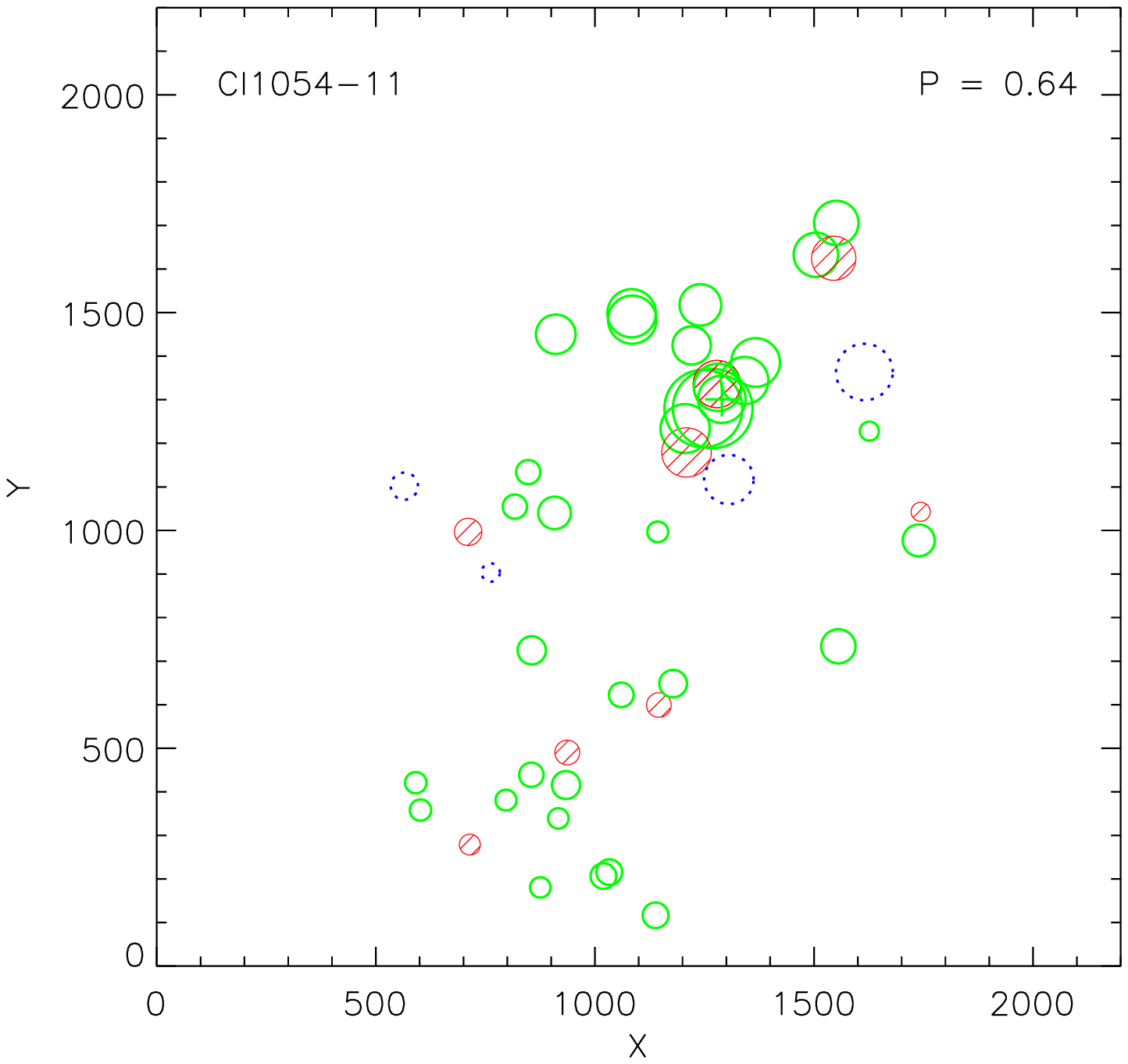}
   \includegraphics[width=7.5cm]{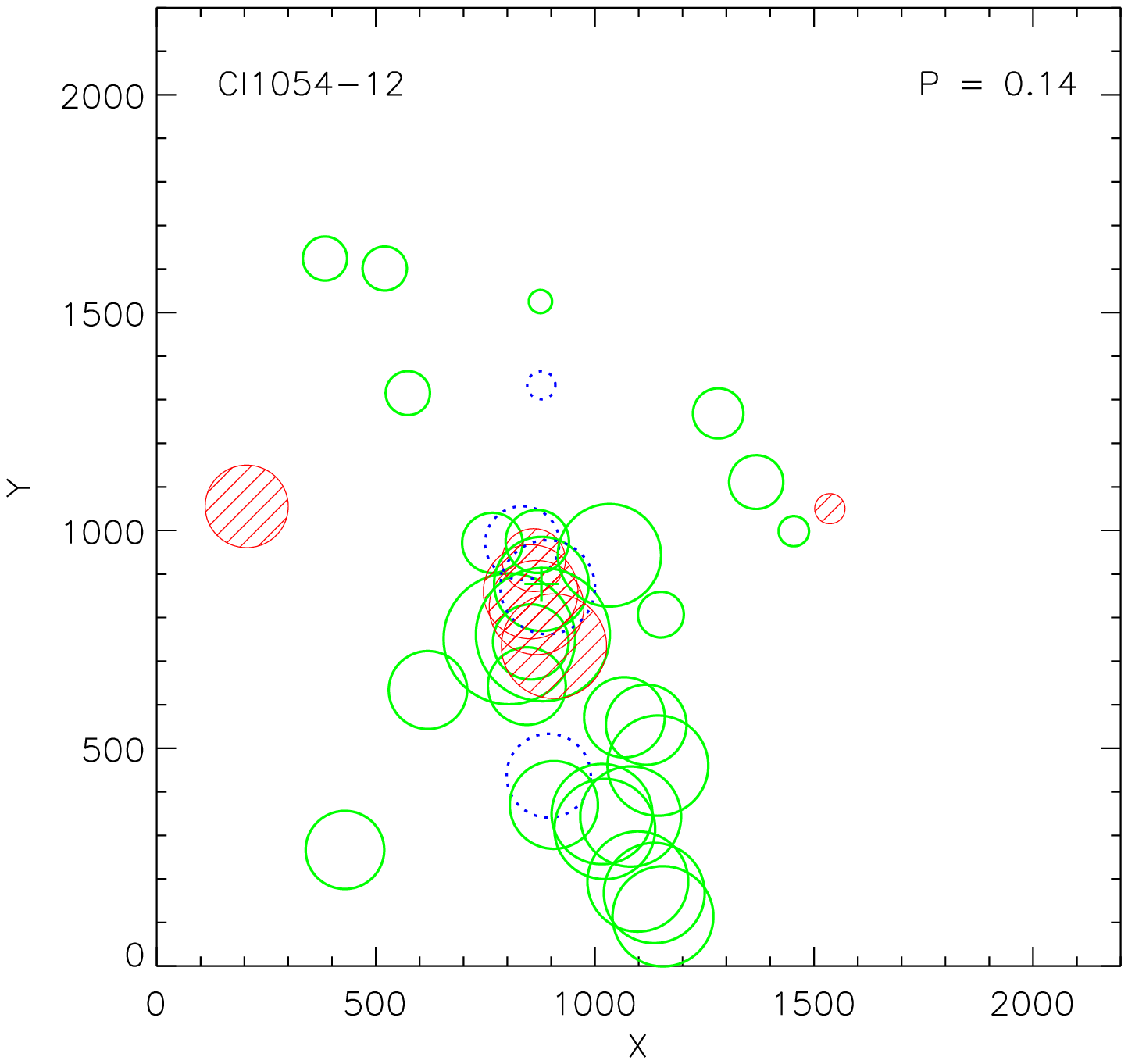}
   \includegraphics[width=7.5cm]{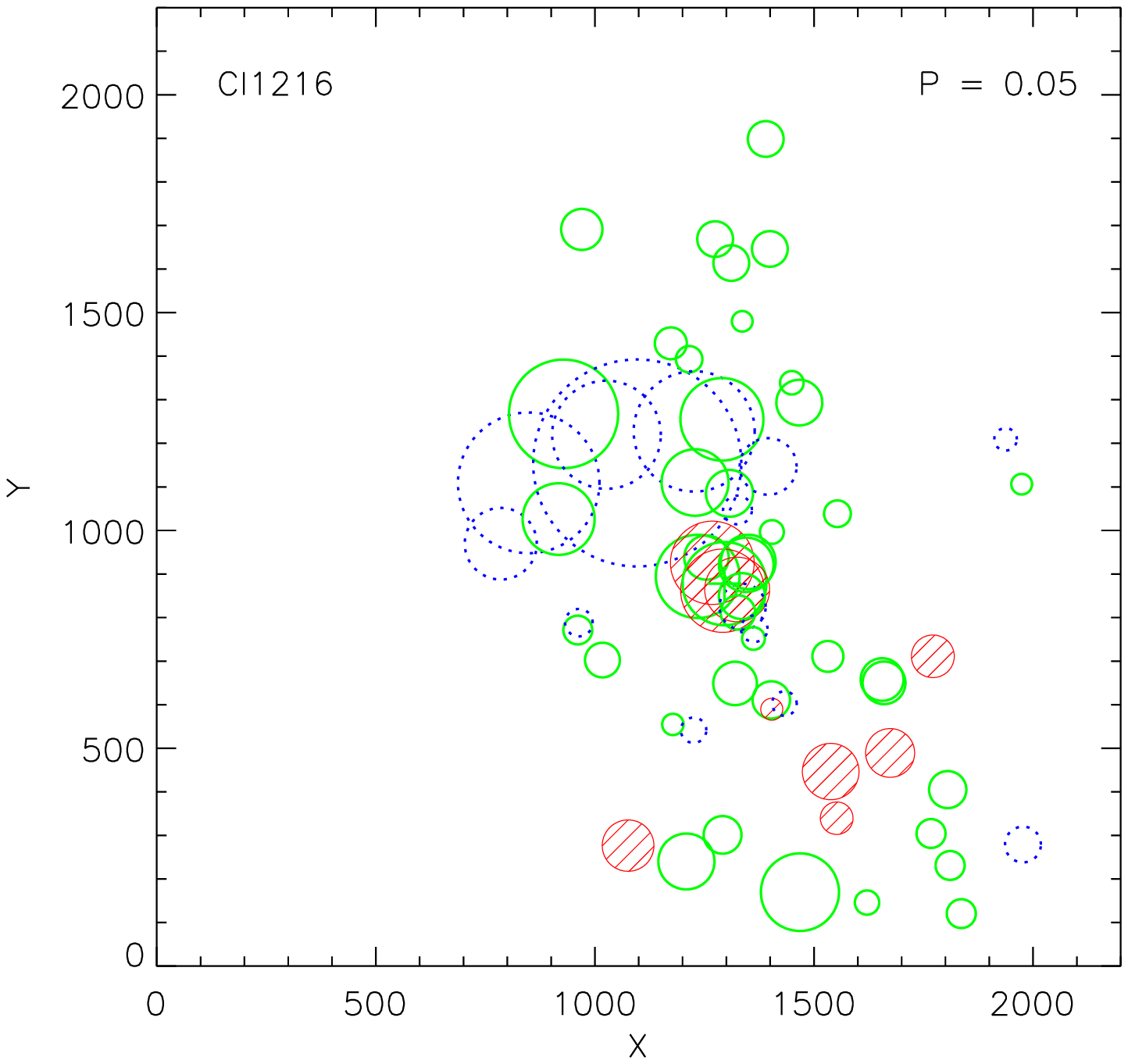}
   \includegraphics[width=7.5cm]{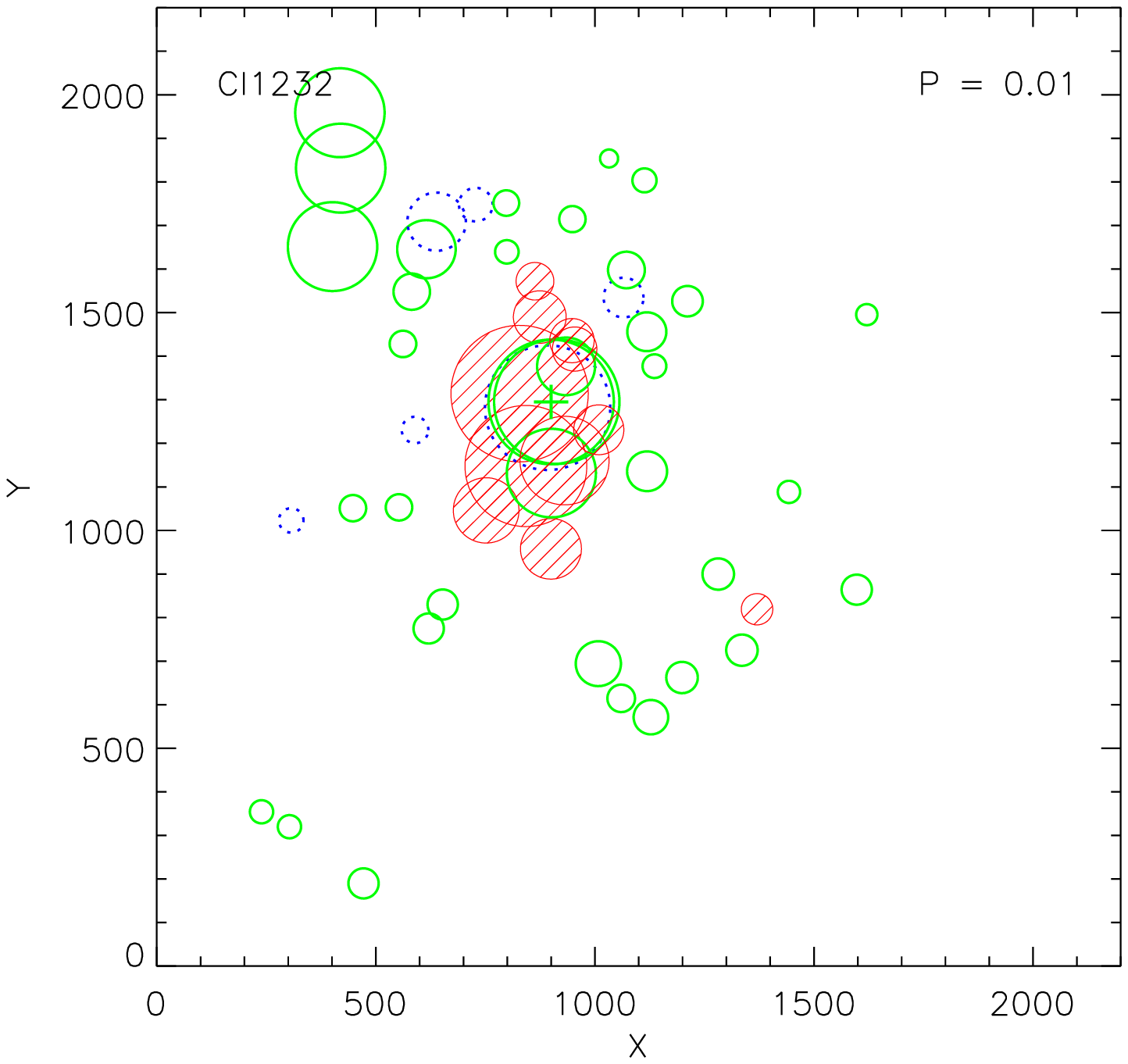}

   \caption{Dressler-Shectman (DS) analysis. The radii of the plotted
   circles is equal to $e^{\frac{\delta}{2}}$ where $\delta$ is the DS
   measurement of local deviation from the global velocity dispersion
   and mean recession velocity. Green circles marked by a solid line
   indicate galaxies with redshift $z_{i}$ such that ($z_{cluster} -
   z_{\sigma}$) $\le$ $z_{i}$ $\le$ ($z_{cluster} + z_{\sigma}$), red
   circles indicated by hashed areas, galaxies with $z_{i}$ $>$
   ($z_{cluster} + z_{\sigma}$), and blue circles indicated by a
   dotted line, $z_{i}$ $<$ ($z_{cluster} - \sigma_{cluster}$). The
   measure of substructuring significance $P$ is provided at the top
   right hand corner of plots for each cluster.\label{fig:DShigh}}
\end{figure*}

More robust evidence for substructure requires its significance and
its properties to be quantified. \citet{Dress_Shect:1988} devised a
statistical test that uses galaxy velocities and positions to
constrain cluster substructure in three dimensions. The principle of
the method is the following. Starting from a list of cluster members
with measured positions and velocities, for each galaxy one finds the
ten listed nearest neighbours on the sky. The local mean velocity and
velocity dispersion are computed from this sample of $11$ galaxies.
These quantities, defined for each galaxy in the list, are then
compared to the global cluster mean and velocity dispersion using the
parameter $\delta$ defined to be:

\begin{equation}
\delta^2 = \frac{11}{\sigma^2}\times\left[\left(\bar{v}_{\rm
      local}-\bar{v}\right)^2+\left(\sigma_{\rm local}-\sigma\right)^2\right]
\end{equation}

\noindent where $\sigma$ and $\bar{v}$ are the cluster velocity
dispersion and the cluster redshift respectively, as given for each
cluster in Section \ref{cvddesc}, while $\sigma_{\rm local}$ and
$\bar{v}_{\rm local}$ are the quantities defined locally for each
galaxy.

\citet{Dress_Shect:1988} define the cumulative deviation $\Delta$ as
the sum of the $\delta$ for all the cluster members $N_{\rm g}$. Note
that the $\Delta$ statistic is similar to a $\chi^2$: if the cluster
velocity distribution is close to Gaussian and local variations
are just sampling fluctuations, then $\Delta$ will be of the order of
$N_{\rm g}$.

This technique does not allow a direct identification of galaxies
belonging to a detected substructure. The position of substructures
can however be identified by plotting the distribution of the galaxies
on the sky using symbols whose size is proportional to the parameter
${\delta}$, thus quantifying the local deviation from the global
kinematics of the cluster. Such an analysis is shown for our 5
clusters in Figure
\ref{fig:DShigh}. In each panel, the size of the symbols is
proportional to $e^{\delta/2}$ and a cross symbol marks the position
of the BCG. The circle symbols are coloured according to the criteria
described in Section \ref{desxyplots}.
To first order the presence of many large circles in a given area
should correspond to substructure, i.e. a correlation both spatially
and kinematically of separate structures. Our plots in Figure
\ref{fig:DShigh} indicate that complex dynamical structure may be
present for all clusters although significant substructuring is
detected for only Cl\,1216 and Cl\,1232.

The $\Delta$ statistic can be used to give a quantitative estimate of
the significance of the putative substructure. We have computed a set
of $1000$ Monte Carlo simulations for each cluster by randomly
re-shuffling the velocities of the cluster members. This removes any
significant 3D substructure. The statistical significance of the
observed substructure can then be quantified by the fraction $P$
\citep{Dress_Shect:1988} of the simulations that yield $\Delta$ values
larger than the observed one where a {\it small} value of $P$
corresponds to a {\it high} significance. In Table \ref{tab:DS} we
list the number of spectroscopic members, the measured $\Delta$
statistic, and the measure of substructuring significance $P$.

\begin{table}
  \begin{center} \begin{tabular}{llll} \hline Cluster name & $N_{\rm
  g}$ & $\Delta$ & $P$ \\ \hline Cl\,1040 & $30$ & $55.11$ & $0.08$ \\
  Cl\,1054-11 & $48$ & $71.77$ & $0.64$ \\ Cl\,1054-12 & $37$ &
  $108.87$ & $0.14$ \\ Cl\,1216 & $67$ & $135.58$ & $0.05$ \\ Cl\,1232
  & $54$ & $96.57$ & $0.01$ \\ \hline \end{tabular} \end{center}
  \caption{Number of spectroscopic members ($N_{\rm g}$), value of the
  $\Delta$ statistic, and the measure of substructuring significance
  $P$, for each of the clusters used in the present analysis (see text
  for details).} \label{tab:DS}
\end{table}


For two of our clusters (Cl~1232 and Cl~1216), substructure is
detected by the DS test with more than 95\% confidence. This agrees
with the spatial segregation of galaxies in different velocity
intervals seen in Figure \ref{fig:xyplots} and with the existence of
peaks redward (Cl~1232) or blueward (Cl~1216) of the cluster redshift
in the velocity histograms (Figure \ref{fig:velhistohigh}). In Cl~1232
the red-shifted galaxies are mainly projected near the cluster centre,
while in Cl~1216 the blue-shifted galaxies are offset North-East of
the BCG. In two clusters (Cl~1040 and Cl~1054-12), substructure
appears to be present, based on the DS probability, but the available
data do not yet provide firm statistical evidence. No evidence for
substructure is found in Cl~1054-11. Interestingly, the two clusters
with the clearest substructure are the ones with the largest velocity
dispersions, both exceeding 1000 $\,{\rm km}\,{\rm s}^{-1}$. At these
redshifts, equilibrium clusters with such high velocity dispersions
are rare according to the standard hierarchical $\Lambda$CDM scenario
(e.g. \citealt{Eke_etal:1996}). The substructure analysis suggests
that both clusters have considerable depth in redshift, indicating
that they may still be collapsing along or close to the
line-of-sight. One should therefore be extremely cautious when using
their measured velocity dispersions to estimate their masses. Such
non-virialised structures are not unexpected in $\Lambda$CDM models
\citep{Evrard_etal:2002}.

\section{Summary}\label{summary}

We present spectroscopic observations of 5 clusters, 4 at redshift of
0.7--0.8, 1 at redshift $\sim$ 0.5, completed as part of the ESO
Distant Cluster Survey (EDisCS). We demonstrate how our target
selection strategy has maximised our cluster member dataset. We
present spectroscopic catalogues of positions, redshifts and I-band
magnitudes for galaxies observed in the fields of our clusters. We
have acquired data for 236 cluster members and the number of cluster
members per cluster ranges between 30 and 67. Our spectroscopic
success rate at targeting members was 30--75\%, averaging 50\%. Our
cluster velocity dispersions measured using a robust biweight
estimator range from 400 to 1100 $\,{\rm km}\,{\rm s}^{-1}$. We find
significant evidence for substructure in two clusters, one at $z \sim
0.79$ and the other at $z \sim 0.54$. Both clusters have velocity
dispersions exceeding 1000 $\,{\rm km}\,{\rm s}^{-1}$ and are unlikely
to be fully virialised: their velocity dispersions may therefore not
provide a robust estimate of their mass. The properties of our 5
EDisCS clusters cover a range of velocity dispersions, richness and
substructuring typical of our survey. Our full spectroscopic dataset
will allow a comprehensive study of galaxy evolution as a function of
redshift and cluster environment.

\begin{acknowledgements}
We thank Timothy Beers, Andrea Biviano and Alan Dressler for helpful
discussions. We thank the staff of the ESO Very Large Telescope
observatory, Paranal, Chile, for their help and support before, during
and after observations. The anonymous referee is gratefully thanked
for a constructive report which improved the quality of the final
manuscript. C.~H. acknowledges the use of the program ROSTAT. C.~H.
and B~.M.~P. gratefully acknowledge the grant support of the Italian
National Research Centre, grant number CNRG008871, and a grant from
the "Fondo per gli Investimenti della Ricerca di Base" of the Italian
Ministery of Education, University and Research (grant
RBAU018Y7E). G.~D.~L. thanks the Alexander von Humboldt Foundation,
the Federal Ministry of Education and Research, and the Programme for
Investment in the Future (ZIP) of the German Government for financial
support. R.~P.~S. and G.~R. acknowledge the support by the Deutsche
Forschunggemeinschaft (DFG), SFB 375 (Astroteilchenphysik).

\end{acknowledgements}

\bibliographystyle{aa}
\bibliography{AA20041304}

\appendix

\section{Cosmic ray removal algorithm}\label{cosmic}

For the `long masks' typically 4 or 8 exposures of 30$\,$minutes were
available and the following method was used
(\citealt{Milvang-Jensen:2003}; Milvang-Jensen \&
Arag{\'o}n-Salamanca, in prep.).
Consider the example of 8 exposures. To be able to flag deviating
values among the 8 values available at a given pixel position we must
calculate the expected standard deviation (hereinafter `sigma') in the
absence of cosmics. One estimate of sigma can be computed from the
median using the CCD noise model (photon noise and read-out
noise). Since the numerous sky lines have large intrinsic intensity
variations this sigma can however significantly underestimate the
actual frame-to-frame variation. The sigma must therefore be measured
from the data.
The number of pixels affected by cosmics in a given 30$\,$minute
exposure is approximately 0.2\%. It follows that for a given pixel
position at most 2 values are expected to be affected by cosmics, and
thus the lowest 6 values can be used to compute a sigma. As a
statistic we used the RMS (run~2) or the MAD (mean absolute deviation
from the sample mean) (run~3); the results were found to be similar.
This sigma estimate is biased low with respect to the intrinsic value
(still in the absence of cosmics). The exact factor was found from
simulations using a normal parent distribution. The estimates were
then multiplied by this factor to give a `measured' sigma estimate,
e.g.\ $\sigma_{\rm meas.}
\equiv 1.79 \cdot {\rm MAD}({\rm lowest\,6\,of\,8})$. This estimate
incorporates the large variations due to the varying strength of the
sky lines, while giving a value close to that predicted by the CCD
noise model in regions of the spectra not affected by sky
lines. However, due to small number statistics ($\sim$$\sqrt{6}$
uncertainty) the measured sigma can occasionally be too
small. Therefore, the final sigma was taken to be the maximum of the
measured sigma and the CCD noise model sigma.
Cosmics in the individual frames could now be identified as values
exceeding the median image by more than $N$ sigma. The factor $N$ was
chosen manually on a frame by frame basis in such a way that very few
good (i.e.\ non-cosmic) pixels were flagged. Usually $N$ = 4--5
worked. If however a series of 8 (say) mask exposures was started at
the beginning of the night, sky lines in the first exposures could be
much brighter than in the remainder in a way not predictable using the
measured sigma: in this case a larger threshold $N$ was required.
For run~2 the cosmic maps were used to clean the individual exposures
using linear interpolation within each frame. Following this the
exposures were combined (i.e.\ averaged). For run~3 the cosmic masks
were used to exclude cosmics when combining the frames. It should be
noted that the stability of FORS2 is excellent and the shift in
wavelength and spatial direction during 4 hours (8 exposures) of
uninterrupted observations is small ($\lesssim0.5\,$pixel), allowing
frames to be combined without shifts.

\end{document}